\documentclass[a4paper,12pt]{article}
\usepackage{pdflscape}
\usepackage{amsmath}
\usepackage{cite}
\usepackage{amssymb}
\usepackage{dcolumn}
\usepackage{bm}
\usepackage{color}
\usepackage{epsfig}
\usepackage{amsfonts}
\usepackage{graphicx}
\usepackage{subfigure}
\usepackage{dcolumn}
\usepackage{indentfirst}
\usepackage{authblk}
\usepackage{slashed}
\usepackage{appendix}

\usepackage{booktabs}
\AtBeginDocument{
\heavyrulewidth=.08em
\lightrulewidth=.05em
\cmidrulewidth=.03em
\belowrulesep=.65ex
\belowbottomsep=0pt
\aboverulesep=.4ex
\abovetopsep=0pt
\cmidrulesep=\doublerulesep
\cmidrulekern=.5em
\defaultaddspace=.5em
}

\usepackage[colorlinks=false]{hyperref}
\hypersetup{citecolor=Blue}

\begin{document}

\vspace{80pt}

\centerline{\LARGE  Holographic Schwinger effect with Translational Symmetry Breaking}
\vspace{40pt}
\centerline{
 Sara Tahery,$^{\ast}$
\let\thefootnote\relax\footnote{$^{\ast}$saratahery@htu.edu.cn}
Wenxing Cheng,$^{\dagger}$
\let\thefootnote\relax\footnote{$^{\dagger}$ chengwenxing@cug.edu.cn}
 Zi-qiang Zhang,$^{\dagger\dagger}$
\let\thefootnote\relax\footnote{$^{\dagger\dagger}$ zhangzq@cug.edu.cn} 
and Qin Chang$^{\dagger\dagger\dagger}$
\let\thefootnote\relax\footnote{$^{\dagger\dagger\dagger}$ changqin@htu.edu.cn} }
\vspace{30pt}
{\centerline {$^{\ast,\dagger\dagger\dagger }${\it Institute of Particle and Nuclear physics,  
Henan Normal University,  Xinxiang 453007, China
}}
\vspace{4pt} 
{\centerline {$^{\dagger,\dagger\dagger }${\it School of Mathematics and Physics, China University of
Geosciences, Wuhan 430074, China
}}
\vspace{40pt}

\begin{abstract}
We investigate the holographic Schwinger effect in a background with translational symmetry breaking (TSB) at finite chemical potential. The gravitational background is characterized by two independent parameters: the TSB parameter \(\alpha\), which controls momentum relaxation, and the chemical potential \(\mu\), which determines the finite density of the dual field theory. Using the potential analysis method, we derive the total potential governing the pair production process and examine its dependence on \(\alpha\), \(\mu\), the external magnetic field, and the ratio \(\beta=E/E_c\).
Our results show that the effects of \(\alpha\) and \(\mu\) on the Schwinger process strongly depend on the dynamical regime. In the subcritical regime, increasing either \(\alpha\) or \(\mu\) lowers the potential barrier and facilitates pair production. However, near and above the critical electric field, the roles of these two parameters become qualitatively different. While increasing the chemical potential lowers the total potential and enhances the Schwinger pair production process, increasing the translational symmetry breaking parameter shifts the potential upward and suppresses the production process.
We further show that, at fixed $\beta=E/E_c$, a perpendicular external magnetic field lowers the effective potential barrier and thereby facilitates the Schwinger process, while the physical electric field changes accordingly through the magnetic-field dependence of $E_c$. The corresponding pair production rate is not independently calculated; instead, its qualitative behavior is characterized through a proxy derived from the total potential.
Overall, our analysis provides a comprehensive picture of how translational symmetry breaking, finite density, and external magnetic fields influence holographic non-perturbative pair production.
\end{abstract}


\tableofcontents
\section{Introduction}
An intriguing non-perturbative phenomenon that naturally lends itself to holographic modeling is the Schwinger effect, namely the spontaneous production of particle--antiparticle pairs in the presence of a sufficiently strong external electric field. Originally studied in the context of quantum electrodynamics (QED), the Schwinger effect has attracted considerable attention in holography as a powerful probe of vacuum instability in strongly coupled systems \cite{Semenoff2011,Hou2018,Zhang2016,Sato2013}. In the holographic framework, pair production is described through the dynamics of fundamental strings or probe branes embedded in the bulk geometry, allowing one to determine the critical electric field and analyze the corresponding potential barrier associated with the pair production process. Various extensions of the holographic Schwinger effect have been investigated in the literature, including finite temperature, confinement, chemical potential, and external field effects \cite{Zi-qiang2018,zhu2021,Kazem1504}, providing a rich setting for studying non-perturbative vacuum dynamics.

To investigate the influence of translational symmetry breaking (TSB) on the Schwinger effect, one considers holographic models in which momentum conservation is violated in the dual boundary theory. In many holographic constructions, translational symmetry is broken while the background geometry remains spatially homogeneous. Such symmetry breaking can be realized through different bulk mechanisms, including massive gravity models and couplings to additional bulk fields. For example, Vegh's model introduces mass terms for the graviton, explicitly breaking diffeomorphism invariance in the bulk and consequently momentum conservation in the boundary theory \cite{Vegh2013,davison2013}. Another well-known realization is the model of Andrade and Withers, where massless scalar fields with linear spatial profiles provide a controlled mechanism for momentum relaxation \cite{Andrade2013}. Related realizations based on massless two-form fields were also studied in \cite{Groz2018}. A broader and more systematic discussion of translational symmetry breaking mechanisms in holography can be found in \cite{1904.05785}, where both explicit and spontaneous symmetry breaking are analyzed within the framework of massive gravity. Such holographic constructions provide a flexible framework for investigating the role of translational symmetry breaking in strongly coupled systems and its effects on various physical observables, including drag force \cite{ours2406} and non-perturbative pair production processes.

The role of external magnetic fields in the holographic Schwinger effect has also been extensively studied. Previous investigations demonstrated that magnetic fields can significantly modify the pair production process depending on their orientation relative to the electric field \cite{satoyo1303,Bolo1210,Koji1403}. Beyond the simplest AdS backgrounds, the holographic Schwinger effect has been explored in finite-density systems and confining geometries \cite{Zhang2016,Hou2018}, higher-derivative theories such as Gauss--Bonnet gravity \cite{Zi-qiang2018}, and QCD-inspired holographic models \cite{zhu2021}. More recently, magnetized holographic backgrounds \cite{zhou1912}, flavor-dependent configurations \cite{Lin2025}, and anisotropic geometries \cite{Wen2506} have been investigated, revealing a rich interplay among background fields, symmetry-breaking mechanisms, and vacuum instability.

Motivated by these developments, in this paper we investigate the holographic Schwinger effect in a background with explicit translational symmetry breaking and finite chemical potential. Our primary goal is to understand how the translational symmetry breaking parameter \(\alpha\) and the chemical potential \(\mu\) influence the total potential, the critical electric field, and the corresponding Schwinger pair production process. Particular attention is devoted to clarifying the distinct physical roles of these two parameters. In our setup, the parameter \(\alpha\) exclusively characterizes translational symmetry breaking and momentum relaxation, whereas the chemical potential \(\mu\) controls the finite density of the dual field theory without introducing any form of symmetry breaking. We further extend the analysis by incorporating an external magnetic field and systematically investigate its influence on the pair production process in different electric field regimes.

The organization of the paper is as follows. In section~\ref{sec:TSB}, we review the holographic background geometry with translational symmetry breaking and finite chemical potential. In section~\ref{sec:SchTSB}, we study the holographic Schwinger effect in this background by computing the total potential for a probe quark--antiquark pair through Wilson loop analysis and determining the corresponding critical electric field. We first investigate the behavior of the total potential in the absence of magnetic field for the subcritical, critical, and supercritical electric field regimes. We then extend the analysis to the case with external magnetic field and examine the corresponding modifications of the Schwinger effect. Finally, we qualitatively analyze the Schwinger pair production rate and discuss its consistency with the behavior of the total potential in section~\ref{se:Gamma}. In section~\ref{Summ}, we summarize the main results and discuss possible future directions.
\section{Translational Symmetry Breaking Background} \label{sec:TSB}
We consider a bulk geometry where translational symmetry is explicitly broken via massless scalar fields. The background metric and field content are given by \cite{Andrade2013},
\begin{equation}\label{eq:metric}
ds^2 = -f(r)dt^2 + \frac{dr^2}{f(r)} + r^2\delta_{ab} dx^a dx^b, \quad A = A_t(r)dt, \quad \psi_I = \alpha_{Ia}x^a,
\end{equation}
where $f(r)$ denotes the blackening function, $A_t(r)$ is the time component of the Maxwell field, and $\psi_I$ are spatially linear scalar fields responsible for breaking translational invariance. Here, $a$ labels the $d-1$ spatial directions $x^a$, $I$ is an internal index labeling the $d-1$ scalar fields, and $\alpha_{Ia}$ are real arbitrary constants. The parameter $\alpha$ controls the strength of momentum relaxation induced by the linear axion fields and thus explicitly breaks translational invariance in the boundary theory.

Throughout this work, $\alpha$ is interpreted solely as a translational symmetry breaking (TSB) parameter and should not be confused with random disorder. Therefore, increasing $\alpha$ corresponds to stronger momentum dissipation in the dual field theory while preserving the homogeneity of the bulk geometry. It is important to emphasize that the present model remains homogeneous and isotropic despite the presence of $\alpha$, and thus does not represent disorder in the usual condensed matter sense.

To ensure isotropy, the coefficients $\alpha_{Ia}$ are chosen such that,
\begin{align}
\vec{\alpha}_a \cdot \vec{\alpha}_b &= \alpha^2 \delta_{ab}, \label{eq:alpha_orth} \\
\Rightarrow \quad \alpha^2 &\equiv \frac{1}{d-1}\sum_{a=1}^{d-1} \vec{\alpha}_a \cdot \vec{\alpha}_a, \label{eq:alpha_def}
\end{align}
which guarantees that the spatial directions are treated on equal footing.

The gauge field and metric function take the form,
\begin{align}
A_t(r) &= \mu \left(1 - \frac{r_h^{d-2}}{r^{d-2}}\right), \label{eq:At_general} \\
f(r) &= r^2 - \frac{\alpha^2}{2(d-2)} - \frac{m_0}{r^{d-2}} + \frac{\mu^2}{2}\frac{d-2}{d-1} \left(\frac{r_h}{r}\right)^{2(d-2)}, \label{eq:f_general}
\end{align}
where $\mu$ is interpreted as the chemical potential in the dual field theory, and $m_0$ is an integration constant related to the black hole mass.

The constant $m_0$ is fixed by requiring regularity of the metric at the horizon, namely by imposing the condition $f(r_h)=0$. This leads to,
\begin{equation}\label{eq:m0_general}
m_0 = r_h^d \left(1 + \frac{d-2}{2(d-1)}\frac{\mu^2}{r_h^2} - \frac{1}{2(d-2)}\frac{\alpha^2}{r_h^2} \right).
\end{equation}

The Hawking temperature associated with this black hole background is determined from the surface gravity at the horizon and is given by,
\begin{equation}\label{eq:T_general}
T = \frac{f'(r_h)}{4\pi} = \frac{1}{4\pi} \left( d r_h - \frac{\alpha^2}{2r_h} - \frac{(d-2)^2 \mu^2}{2(d-1)r_h} \right),
\end{equation}
which shows that both the chemical potential and the scalar fields lower the temperature through their backreaction on the geometry.

Specializing to the case of a five-dimensional bulk ($d=4$), which corresponds to a four-dimensional dual QFT, we obtain the simplified expressions,
\begin{align}
ds^2 &= -f(r)dt^2 + \frac{dr^2}{f(r)} + r^2 dx_i^2, \quad i=1,2,3, \label{eq:metric_d4} \\
f(r) &= r^2 - \frac{\alpha^2}{4} - \frac{m_0}{r^2} + \frac{\mu^2}{3} \frac{r_h^4}{r^4}, \label{eq:f_d4}
\end{align}
and the constant $m_0$ in Eq.~\eqref{eq:m0_general} becomes,
\begin{equation}\label{eq:m0_d4}
m_0 = r_h^4 \left(1 + \frac{\mu^2}{3r_h^2} - \frac{\alpha^2}{4r_h^2} \right).
\end{equation}

Substituting Eq.~\eqref{eq:m0_d4} into Eq.~\eqref{eq:f_d4} yields the explicit form of the blackening function,
\begin{equation}\label{eq:f_final}
f(r) = r^2\left(1 - \frac{r_h^4}{r^4}\right) - \frac{\alpha^2}{4} \left(1 - \frac{r_h^2}{r^2} \right) - \frac{\mu^2}{3} \frac{r_h^2}{r^2} \left(1 - \frac{r_h^2}{r^2} \right),
\end{equation}
which makes manifest the separate contributions from temperature, scalar fields, and charge.

With Eqs.~\eqref{eq:metric_d4} and \eqref{eq:f_final}, one can identify the following limiting cases:
\begin{itemize}
\item 
For $\alpha = 0$ and $\mu\neq 0$, the resulting geometry reduces to the AdS--Reissner--Nordström planar black brane.
\item
For $\alpha \neq 0$ and $\mu= 0$, the resulting geometry corresponds to an AdS black brane with linear scalar fields $\psi_I = \alpha_{Ia}x^a$.
\item
For $\alpha= 0$ and $\mu= 0$, the metric reduces to the AdS--Schwarzschild planar black brane.
\end{itemize}

In this work, $\mu$ and $\alpha$ play distinct physical roles: $\mu$ controls the finite density, while $\alpha$ governs momentum relaxation and translational symmetry breaking.

The corresponding black hole temperature Eq.~\eqref{eq:T_general} for $d=4$ simplifies to,
\begin{equation}\label{eq:T_d4}
T = \frac{1}{4\pi} \left( 4r_h - \frac{\alpha^2}{2r_h} - \frac{2\mu^2}{3r_h} \right),
\end{equation}
which leads to the inequality,
\begin{equation}\label{eq:T_constraint}
24r_h^2 - 3\alpha^2 - 4\mu^2 \geq 0,
\end{equation}
ensuring the physical requirement $T \geq 0$.\\
All numerical parameter choices used throughout this work satisfy the constraint in Eq.~\eqref{eq:T_constraint},
which guarantees non-negative temperature and physical consistency of the background geometry. For all numerical plots, we fix $r_{0}=5$ and $b=0.4$, corresponding to a fixed horizon radius $r_{h}=2$, while the temperature $T$ is determined by Eq. ~\eqref{eq:T_d4}. The results are presented in terms of the dimensionless quantities $\alpha/T$ and $\mu/T$.

The null energy condition (NEC) for this background geometry leads to the constraint,
\begin{equation}\label{eq:NEC}
\alpha^2 + \frac{4r_h^4 \mu^2}{r^4} \geq 0,
\end{equation}
which is always satisfied for non-negative $\alpha^2$ and $\mu^2$.

At extremality, the temperature vanishes, namely $T=0$, yielding,
\begin{equation}\label{eq:extremal_rh}
r_h^2 = \frac{\alpha^2}{8} + \frac{\mu^2}{6}.
\end{equation}

In this limit, the near-horizon geometry becomes $\text{AdS}_2 \times \mathbb{R}^{d-1}$. The curvature radius of the emergent $\text{AdS}_2$ region is~\cite{Andrade2013},
\begin{equation}\label{eq:AdS2_radius}
\ell_{\text{AdS}_2}^2 = \frac{1}{d(d-1)} \cdot \frac{(d-1)\alpha^2 + (d-2)^2 \mu^2}{\alpha^2 + (d-2)^2 \mu^2},
\end{equation}
which smoothly reduces to a finite value even when $\mu = 0$, provided that $\alpha \neq 0$. This confirms that translational symmetry breaking alone can support an $\text{AdS}_2$ throat in the IR. 
\section{Potential analysis with TSB background}\label{sec:SchTSB}
The holographic realization of the Schwinger effect is based on the correspondence between vacuum pair production in the boundary field theory and the dynamics of fundamental strings in the bulk geometry \cite{Maldacena1997}. In this picture, a virtual quark--antiquark pair is represented by the endpoints of an open string attached to a probe D3-brane, while the string worldsheet encodes the effective potential barrier associated with pair production \cite{Semenoff2011}. The pair production rate is governed by the competition between the external electric field, which tends to separate the string endpoints, and the bulk geometry, which resists this separation. The classical configuration of the string profile, extending from the probe brane into the bulk, therefore plays a central role in the holographic analysis. It provides a geometric description of the quark--antiquark system and allows one to compute the total potential governing the pair production process. This framework provides the stage for evaluating the Coulomb potential via the Wilson loop.

Our aim is to evaluate the Coulomb potential through the holographic computation of the rectangular Wilson loop on the probe D3-brane. This is achieved by calculating the classical action of an open string whose endpoints lie on the probe D3-brane \cite{Semenoff2011}, as illustrated in Fig.~\ref{string-profile}. Such a setup provides a direct holographic analogue of the Wilson loop analysis for extracting the quark--antiquark potential in the background geometry.

Consider the background metric Eq.~\eqref{eq:metric_d4} with components,
\begin{equation*}
 G_{tt}(r) = -f(r), \quad 
 G_{xx}(r)=G_{yy}(r)=G_{zz}(r) = r^2, 
 \quad 
 G_{rr}(r) = \frac{1}{f(r)}.
\end{equation*}
We choose the string worldsheet coordinates as \(\sigma^a = (\tau, \sigma)\) and impose the static gauge,
\begin{equation}\label{gauge}
t = \tau, \qquad x = \sigma.
\end{equation}
The radial coordinate of the classical string profile depends only on \(\sigma\),
\begin{equation}\label{eta}
r = r(\sigma).
\end{equation}
The Nambu--Goto Lagrangian density is given by,
\begin{equation}\label{Lgeneral}
\mathcal{L} = \sqrt{-\det \mathcal{G}_{ab}},
\qquad
\mathcal{G}_{ab}
=
\frac{\partial x^\mu}{\partial \sigma^a}
\frac{\partial x^\nu}{\partial \sigma^b}
G_{\mu\nu}.
\end{equation}
Explicitly,
\begin{equation}\label{Lresult}
\mathcal{L}
=
\sqrt{-G_{tt}(G_{xx}+G_{rr}r'^2)}
=
\sqrt{f(r)r^2+r'^2},
\end{equation}
where \(r'=\partial_\sigma r\).
Since the Lagrangian does not explicitly depend on \(\sigma\), the corresponding Hamiltonian density is conserved,
\begin{equation}\label{Lconstant}
\mathcal{L}
-
r'
\frac{\partial \mathcal{L}}{\partial r'}
=
\sqrt{f(r)r^2+r'^2}
-
\frac{r'^2}{\sqrt{f(r)r^2+r'^2}}
=
\text{constant}.
\end{equation}
Applying the boundary condition at the turning point,
\begin{equation}\label{Lbound}
r=r_c,
\qquad
r'=0,
\end{equation}
the constant becomes,
\begin{equation}\label{consterm}
\text{constant}
=
r_c\sqrt{f(r_c)}.
\end{equation}

\begin{figure}[h!]
\begin{center}
\includegraphics[width=10cm]{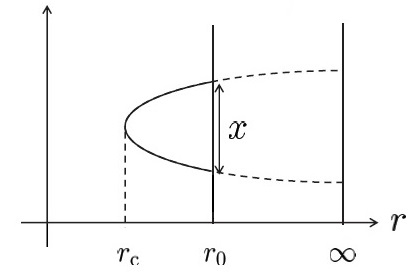}
\end{center}
\caption{String profile in the background geometry.}
\label{string-profile}
\end{figure}
From Eq.~\eqref{Lconstant}, the differential equation for the string profile is obtained as,
\begin{equation}\label{detadx}
\frac{dr}{dx}
=
\sqrt{
f(r)r^2
\left(
\frac{f(r)r^2}{f(r_c)r_c^2}-1
\right)
}.
\end{equation}
Integrating, the separation between the pair endpoints becomes,
\begin{equation}\label{xvfc}
x
=
2
\int_{r_c}^{r_0}
dr
\sqrt{
\frac{f(r_c)r_c^2}
{f(r)r^2\left(f(r)r^2-f(r_c)r_c^2\right)}
}.
\end{equation}
The total potential energy, including the Coulomb potential and static energy, is given by,
\begin{equation}\label{V}
V_{CP+SE}
=
2T_F
\int_{r_c}^{r_0}
dr
\sqrt{
\frac{f(r)r^2}
{f(r)r^2-f(r_c)r_c^2}
}.
\end{equation}
Including the effect of the external electric field,
\begin{equation}\label{vtot}
V_{\text{tot}}
=
V_{CP+SE}-Ex.
\end{equation}
The critical electric field is obtained from the DBI action \cite{sato1309},
\begin{equation}\label{Ecr}
E_{\text{cr}}
=
T_F\sqrt{-G_{tt}(r_0)G_{xx}(r_0)}
=
T_Fr_0\sqrt{f(r_0)}.
\end{equation}
Here, $T_{F}=1/(2\pi\alpha')$ denotes the fundamental string tension, with the numerical value used in all computations given by $T_F=1$.\\
When an external magnetic field is introduced, its effect enters through the DBI action of the probe D3-brane. The background field strength \(F_{\mu\nu}\), containing both electric and magnetic components, modifies the effective string dynamics and consequently changes the total potential governing pair production. This formalism was developed in \cite{satoyo1303,Bolo1210,Koji1403}.

Consider magnetic fields \(B_\parallel\) and \(B_\perp\) parallel and perpendicular to the electric field direction, respectively. The DBI determinant becomes,
\begin{eqnarray}
&-\det
\begin{pmatrix}
G_{tt}(r_0)&-2\pi \alpha 'E &0 &0\\
2\pi \alpha 'E &G_{xx}(r_0) &0 &2\pi \alpha 'B_\perp \\
0 &0&G_{yy}(r_0) &-2\pi \alpha 'B_\parallel\\
0& -2\pi \alpha 'B_\perp&2\pi \alpha 'B_\parallel &G_{zz}(r_0)
\end{pmatrix}
\nonumber\\
\nonumber\\
=&\,
-G_{tt}(r_0)G_{xx}^3(r_0)
-
(2\pi\alpha')^2
G_{tt}(r_0)G_{xx}(r_0)
(B_\parallel^2+B_\perp^2)
-(2\pi\alpha')^2
G_{xx}^2(r_0)E^2
-
(2\pi\alpha')^4E^2B_\parallel^2.\nonumber\\
\end{eqnarray}
The critical electric field is therefore,
\begin{eqnarray}\label{EcrB}
E_{\text{cr}}
&=&
T_F
\sqrt{-G_{tt}(r_0)G_{xx}(r_0)}
\sqrt{
1+\frac{B_\perp^2}
{T_F^2G_{xx}^2(r_0)+B_\parallel^2}
}
\nonumber\\
&=&
T_Fr_0\sqrt{f(r_0)}
\sqrt{
1+\frac{B_\perp^2}
{T_F^2r_0^4+B_\parallel^2}
},
\qquad B\neq0.
\end{eqnarray}
We define the dimensionless ratio,
\begin{equation}\label{alpha}
\beta=\frac{E}{E_{\text{cr}}}.
\end{equation}
Substituting the explicit expressions, the total potential becomes,
\begin{eqnarray}\label{vtotfinal}
V_{\text{tot}}
&=&
2T_F
\int_{r_c}^{r_0}
dr
\Biggl[
\sqrt{
\frac{f(r)r^2}
{f(r)r^2-f(r_c)r_c^2}
}
\nonumber\\
&&
-
\beta r_0\sqrt{f(r_0)}
\sqrt{
1+\frac{B_\perp^2}
{T_F^2r_0^4+B_\parallel^2}
}
\sqrt{
\frac{f(r_c)r_c^2}
{f(r)r^2(f(r)r^2-f(r_c)r_c^2)}
}
\Biggr].
\end{eqnarray}
We now introduce the dimensionless variables,
\begin{equation}\label{aby}
a=\frac{r_c}{r_0},
\qquad
b=\frac{r_h}{r_0},
\qquad
y=\frac{r}{r_c}=\frac{r}{ar_0}.
\end{equation}
Equation~\eqref{vtotfinal} then becomes,
\begin{eqnarray}\label{vtotfinalaby}
V_{\text{tot}}
&=&
2ar_0T_F
\int_{1}^{\frac{1}{a}}
dy
\Biggl(
\sqrt{
\frac{f(y)y^2}
{f(y)y^2-f_c}
}
\nonumber\\
&&
-
\beta r_0\sqrt{f_0}
\sqrt{
1+\frac{B_\perp^2}
{T_F^2r_0^4+B_\parallel^2}
}
\sqrt{
\frac{f_c}
{a^2r_0^2f(y)y^2(f(y)y^2-f_c)}
}
\Biggr),
\end{eqnarray}
where,
\begin{eqnarray}\label{fy}
f(y)
&=&
a^2r_0^2y^2
\left(
1-\frac{b^4}{a^4y^4}
\right)
-
\frac{\alpha^2}{4}
\left(
1-\frac{b^2}{a^2y^2}
\right)
-
\frac{\mu^2}{3}
\frac{b^2}{a^2y^2}
\left(
1-\frac{b^2}{a^2y^2}
\right),
\nonumber\\
\nonumber\\
f_c
&\equiv &f(y=1)=
a^2r_0^2
\left(
1-\frac{b^4}{a^4}
\right)
-
\frac{\alpha^2}{4}
\left(
1-\frac{b^2}{a^2}
\right)
-
\frac{\mu^2}{3}
\frac{b^2}{a^2}
\left(
1-\frac{b^2}{a^2}
\right),
\nonumber\\
\nonumber\\
f_0
&\equiv & f(y=\frac{1}{a})=
r_0^2(1-b^4)
-
\frac{\alpha^2}{4}(1-b^2)
-
\frac{\mu^2}{3}b^2(1-b^2).
\end{eqnarray}
The parameters \(\alpha\) and \(\mu\) enter the total potential through nonlinear combinations appearing in both the numerator and denominator of the square-root structures. Consequently, their effects on the total potential cannot be inferred from a simple sign analysis and must instead be determined numerically.

To verify the consistency of the obtained expression, several limiting cases can be considered,
\begin{itemize}
\item
For \(\alpha\to0\) with \(\mu\neq0\), translational symmetry breaking disappears and the background reduces to the AdS--Reissner--Nordström planar black brane.
\item
For \(\mu\to0\) with \(\alpha\neq0\), the finite-density contribution vanishes and the geometry reduces to the Einstein--axion black brane.
\item
For \(\alpha\to0\) and \(\mu\to0\), the metric reduces to the AdS--Schwarzschild planar black brane.
\item
For \(B\to0\), the magnetic-field contribution disappears and the standard expression without magnetic field is recovered.
\end{itemize}

Using Eq.~\eqref{aby}, the separation between the pair endpoints becomes,
\begin{equation}\label{xintfin}
x
=
2ar_0
\int_{1}^{\frac{1}{a}}
dy
\sqrt{
\frac{f_c}
{a^2r_0^2f(y)y^2(f(y)y^2-f_c)}
}.
\end{equation}

Now consider placing the probe D3-brane at an intermediate radial position \(r=r_0\). The mass of a single quark is identified with the energy of a string stretching from the probe brane to the horizon,
\begin{equation}\label{mass}
m
=
T_F
\int_{r_h}^{r_0}
dr
\sqrt{\det g_{tr}}
=
T_F(r_0-r_h),
\end{equation}
where the induced metric on the string worldsheet is,
\begin{equation}
g_{tr}
=
\mathrm{diag}\left(
G_{tt}(r),
G_{rr}(r)
\right).
\end{equation}

Equation~\eqref{mass} shows that the quark mass is proportional to the separation between the probe brane and the horizon. Therefore, decreasing this distance lowers the energy threshold required for pair production. In terms of the dimensionless parameter \(b=r_h/r_0\), the mass behaves as \(m\propto(1-b)\), indicating that larger values of \(b\) correspond to lighter quark pairs.
\subsection{Vacuum instability in the absence of magnetic Field}

\subsubsection{Subcritical Regime $\beta<1 $}
\begin{figure}[h!]
\begin{minipage}[c]{1\textwidth}
\tiny{(a)}\includegraphics[width=8cm,height=5cm,clip]{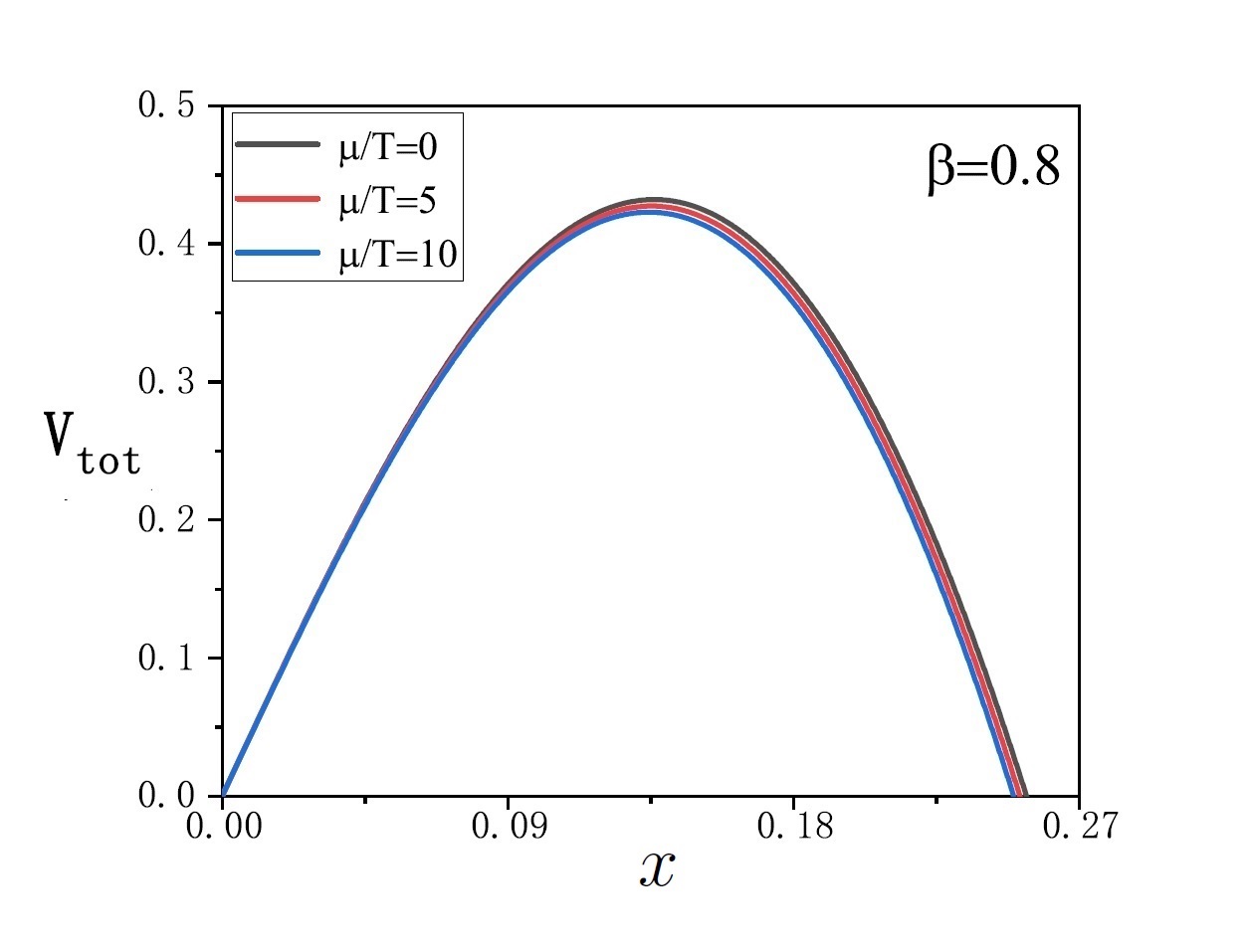}
\tiny{(b)}\includegraphics[width=8cm,height=5cm,clip]{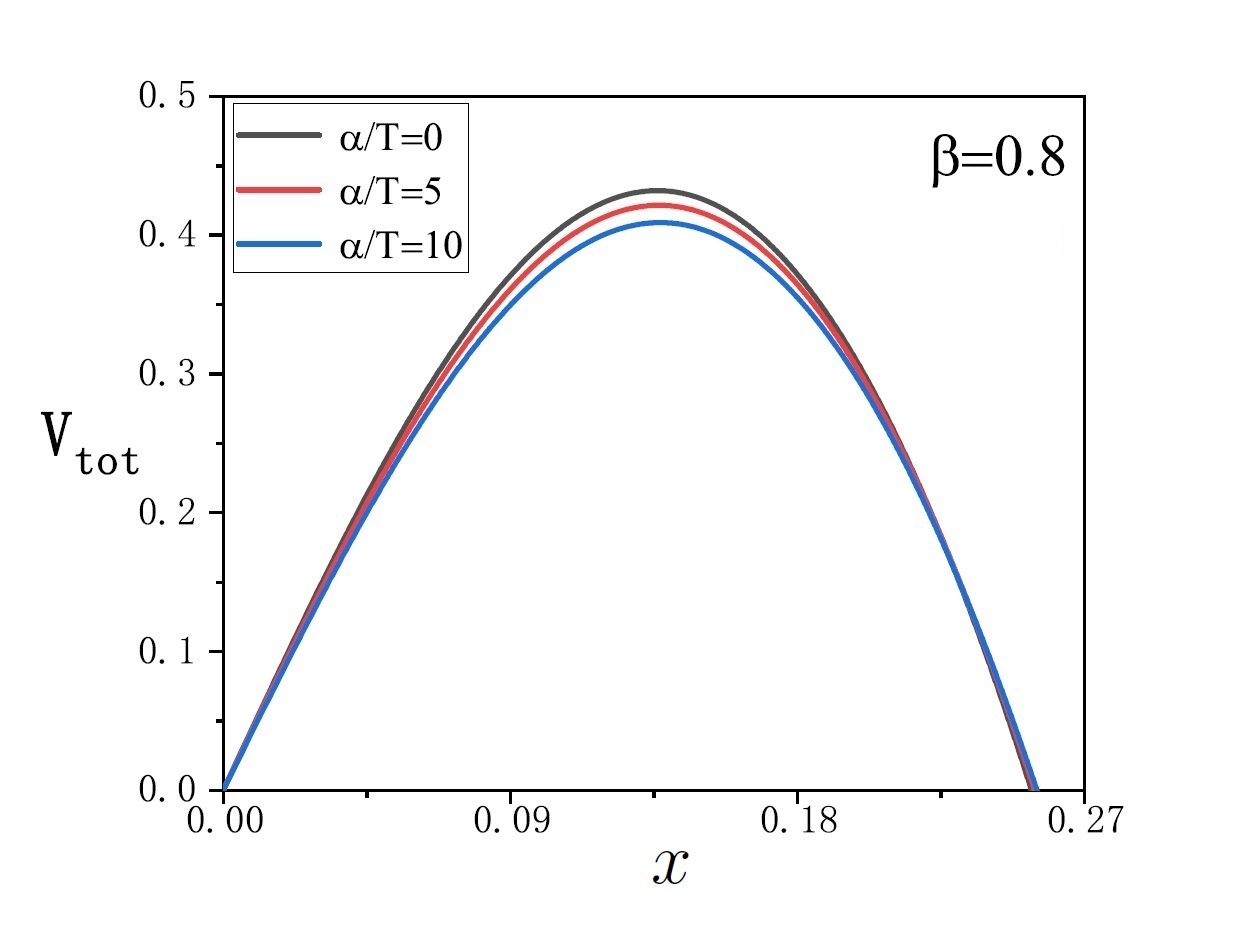}
\end{minipage}
\caption{
Total potential $V_{\mathrm{tot}}$ versus the inter-pair distance $x$ in the absence of magnetic field ($B=0$) for the subcritical regime $\beta=0.8$. The parameters are chosen as $r_0=5$ and $b=0.4$ ($r_h=br_0=2$). The quantities $\alpha$ and $\mu$ are expressed in units of temperature. In panel (a), the momentum relaxation parameter is fixed as $\alpha/T=0$ while $\mu/T$ varies. In panel (b), the chemical potential is fixed as $\mu/T=0$ while $\alpha/T$ varies.} 
\label{tunneling}
\end{figure}
Figure~\ref{tunneling} illustrates the behavior of the total potential in the subcritical regime $(\beta<1)$ without magnetic field. In panel (a), increasing $\mu/T$ slightly lowers the potential barrier and shifts the turning point toward smaller values of $x$. This behavior indicates that finite chemical potential facilitates the tunneling process by weakening the barrier against pair production.

Panel (b) shows the effect of the momentum relaxation parameter. As $\alpha/T$ increases, the height of the potential barrier also decreases and the maximum value of the potential becomes smaller. Therefore, momentum relaxation similarly favors the tunneling process in the subcritical regime.

Comparing both panels, one observes that finite chemical potential and momentum relaxation act in the same direction for $\beta<1$, namely by reducing the potential barrier and enhancing the tunneling probability of virtual pairs.
\subsubsection{Critical Regime $\beta=1$}
 \begin{figure}[h!]
\begin{minipage}[c]{1\textwidth}
\tiny{(a)}\includegraphics[width=8cm,height=5cm,clip]{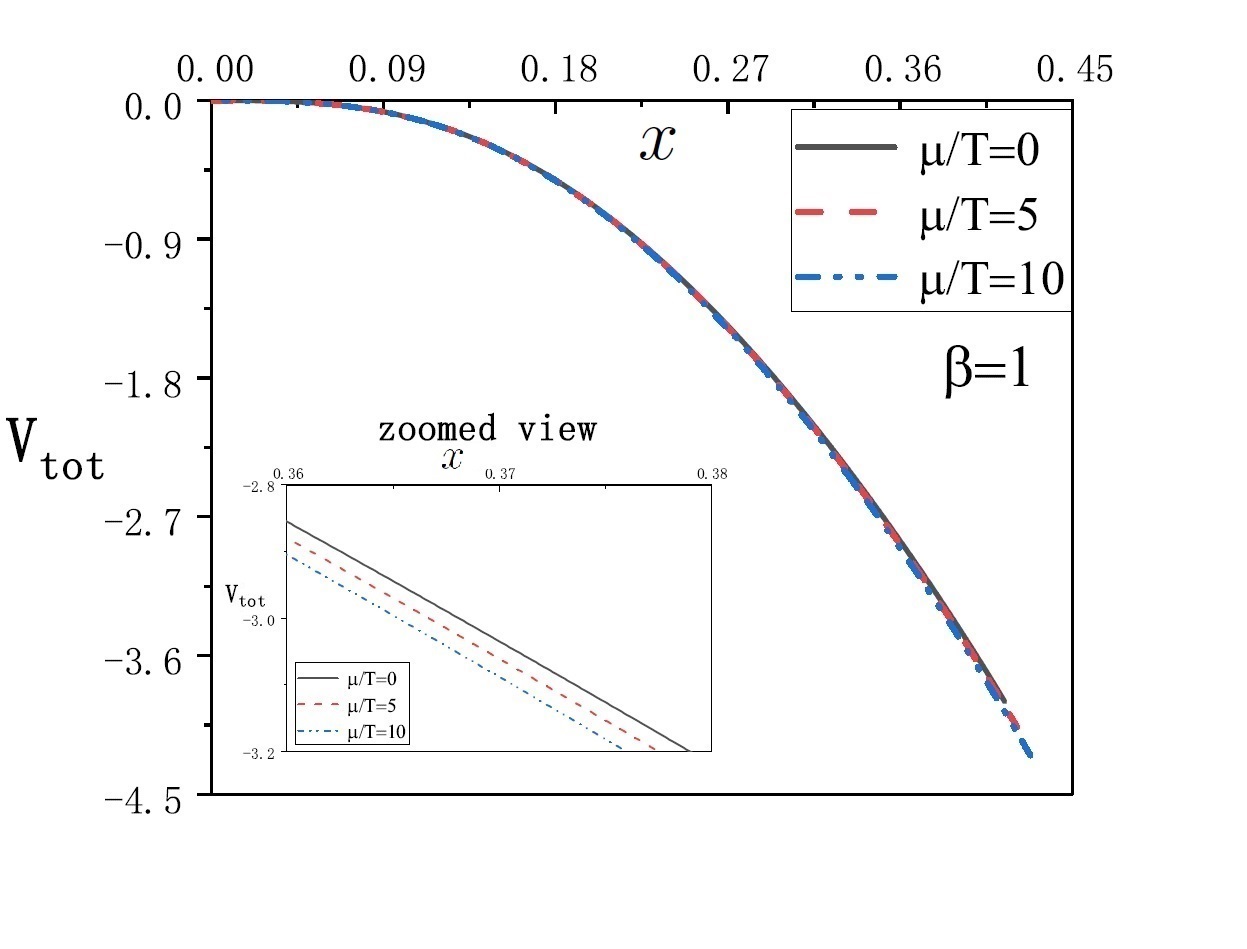}
\hspace{0.2cm}
\tiny{(b)}\includegraphics[width=8cm,height=5cm,clip]{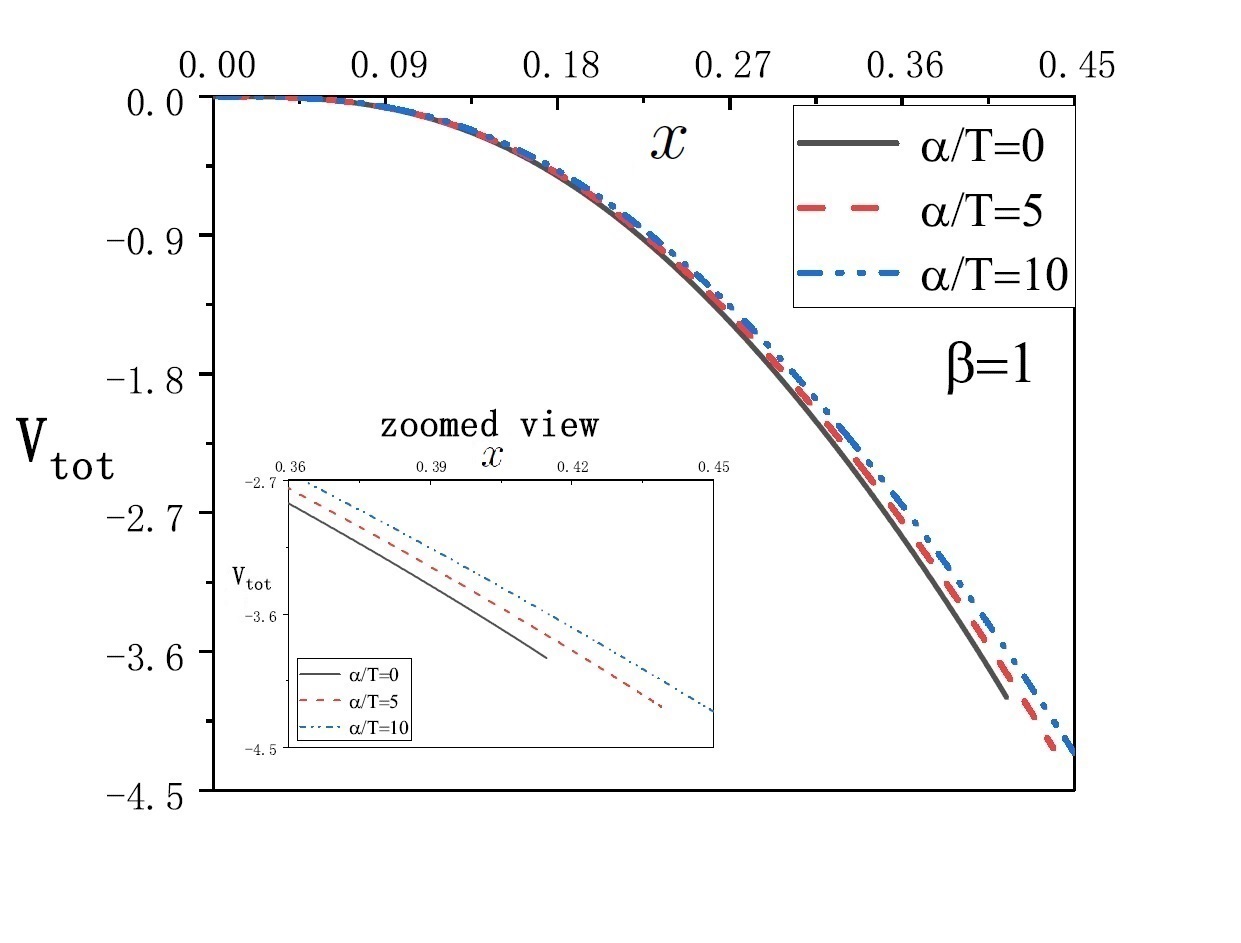}
\end{minipage}
\caption{Total potential $V_{\mathrm{tot}}$ versus the inter-pair distance $x$ in the absence of magnetic field ($B=0$) for the critical regime $\beta=1$. The parameters are fixed as $r_0=5$ and $b=0.4$ ($r_h=2$). In panel (a), $\alpha/T=0$ and $\mu/T$ varies, while in panel (b), $\mu/T=0$ and $\alpha/T$ varies.}
\label{crit}
\end{figure}
Figure~\ref{crit} presents the total potential at the critical electric field $(\beta=1)$ in the absence of magnetic field. In panel (a), increasing $\mu/T$ shifts the potential toward lower values, indicating that finite chemical potential supports the onset of Schwinger pair production near the critical point.

In contrast, panel (b) exhibits the opposite behavior for the momentum relaxation parameter. As $\alpha/T$ increases, the potential curves shift upward, as can be seen more clearly in the zoomed region. This behavior implies that momentum relaxation suppresses Schwinger pair production in the critical regime.

Therefore, unlike the subcritical phase, $\mu/T$ and $\alpha/T$ play opposite roles near the critical electric field. Finite chemical potential favors pair production, whereas stronger momentum relaxation tends to suppress it. This behavior indicates that the role of momentum relaxation changes as the system moves from the subcritical phase to the critical regime.

\subsubsection{Supercritical Regime $\beta>1 $}
 \begin{figure}[h!]
\begin{minipage}[c]{1\textwidth}
\tiny{(a)}\includegraphics[width=8cm,height=5cm,clip]{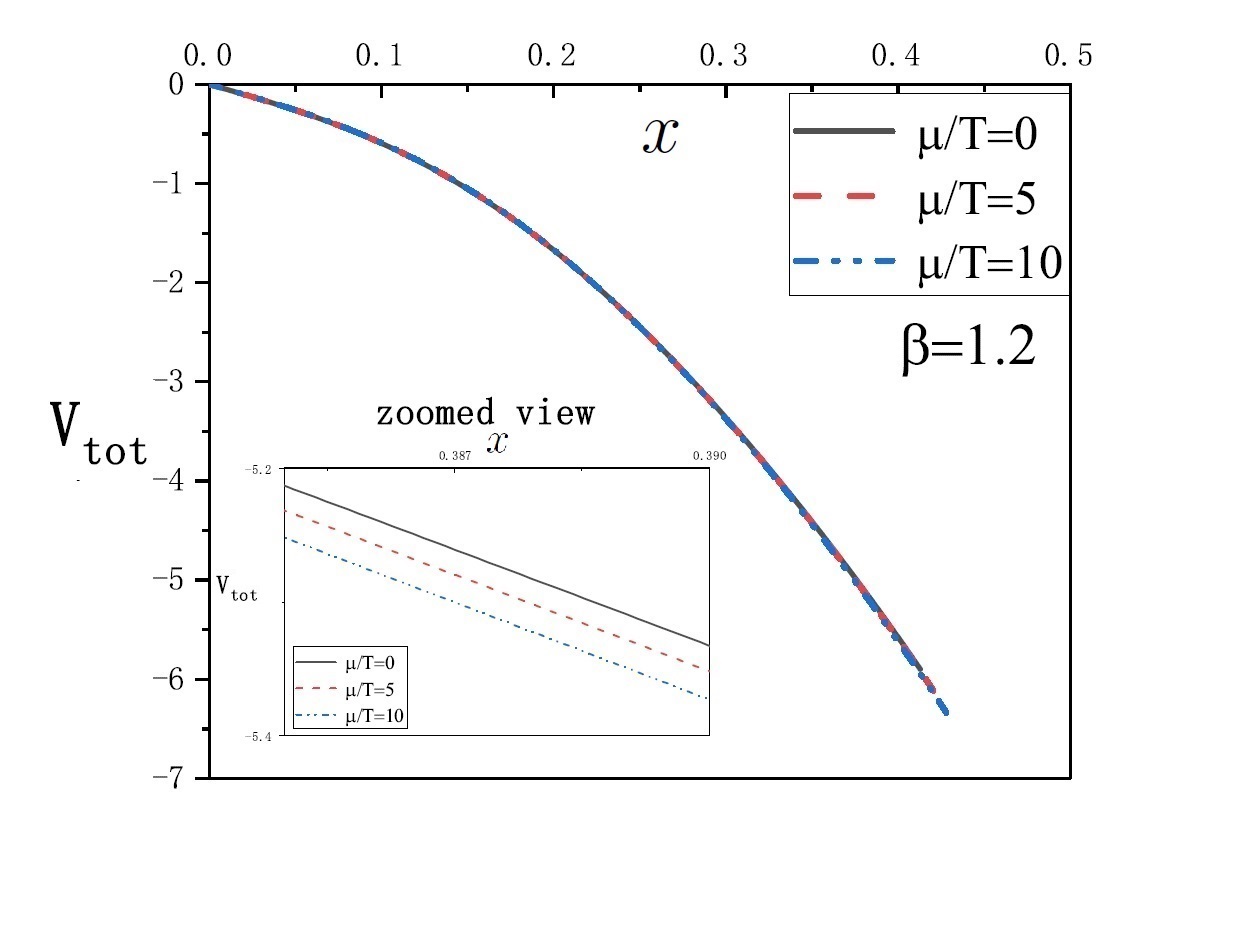}
\hspace{0.2cm}
\tiny{(b)}\includegraphics[width=8cm,height=5cm,clip]{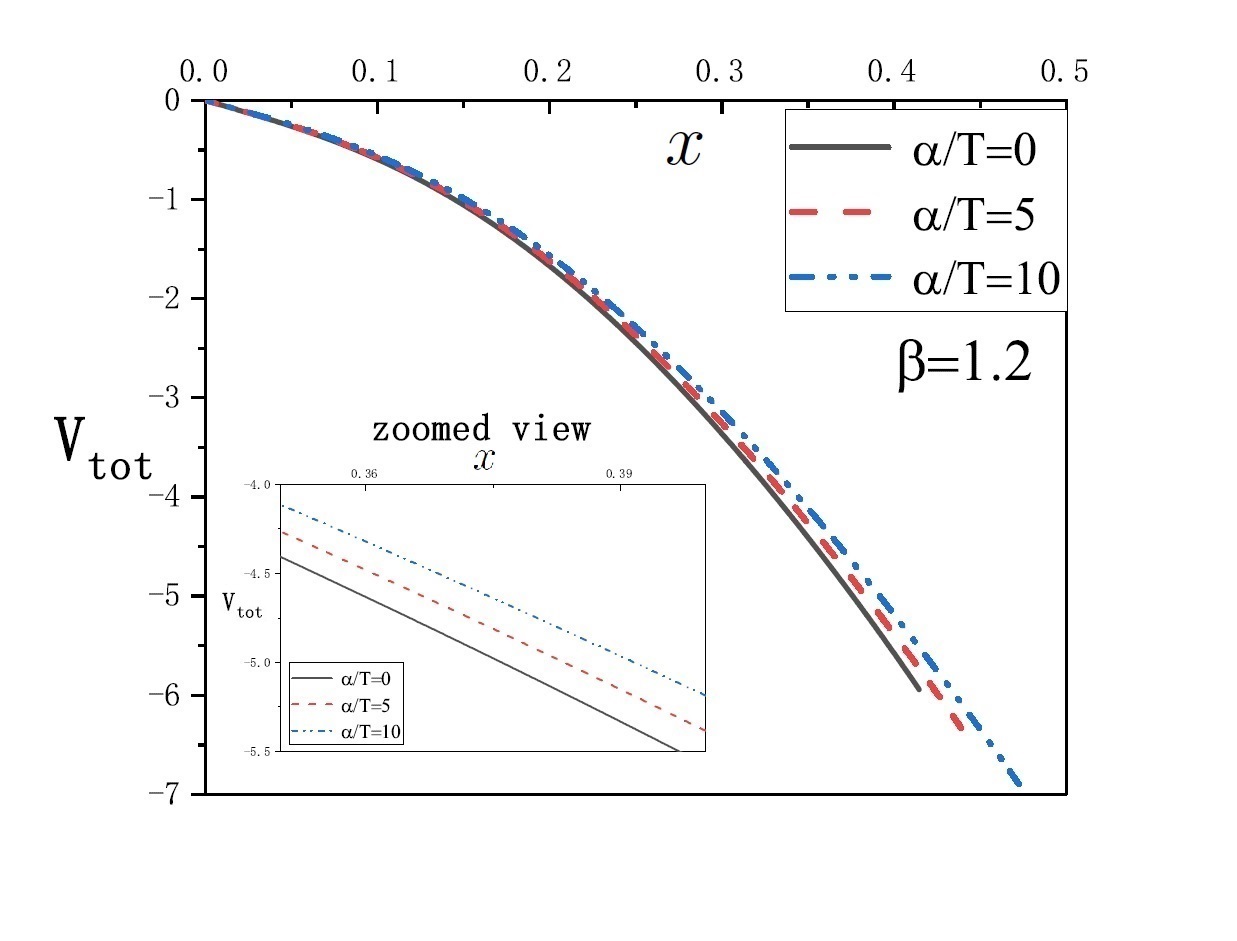}
\end{minipage}
\caption{Total potential $V_{\mathrm{tot}}$ versus the inter-pair distance $x$ in the absence of magnetic field ($B=0$) for the supercritical regime $\beta=1.2$. The parameters are chosen as $r_0=5$ and $b=0.4$ ($r_h=2$). In panel (a), $\alpha/T=0$ and $\mu/T$ varies, while in panel (b), $\mu/T=0$ and $\alpha/T$ varies.}
\label{supercrit}
\end{figure}
Figure~\ref{supercrit} shows the behavior of the total potential in the supercritical regime $(\beta>1)$ without magnetic field. In panel (a), increasing $\mu/T$ shifts the potential toward lower values and strengthens the instability associated with Schwinger pair production.

On the other hand, panel (b) demonstrates that increasing $\alpha/T$ shifts the potential upward. This effect is more visible in the zoomed region, where larger momentum relaxation produces a less negative potential profile. Consequently, momentum relaxation suppresses pair production in the supercritical regime.

Comparing panels (a) and (b), one finds that chemical potential and momentum relaxation produce opposite effects in both the critical and supercritical regimes. Finite chemical potential favors pair production, whereas momentum relaxation suppresses it. This behavior further confirms that the influence of momentum relaxation changes qualitatively as the system crosses the critical electric field.

These findings resonate with conclusions drawn in \cite{Cao2015}, where the Schwinger mechanism in pure electric fields was shown to be sensitive to chiral and background dynamics. Likewise, the recent study \cite{Lin2025} highlights the role of flavor-dependent sources in further amplifying the pair production rate under strong fields.

The present results further indicate that translational symmetry breaking modifies the holographic vacuum structure in a nontrivial way. In the subcritical regime, increasing $\alpha$ lowers the potential barrier and facilitates tunneling. However, near and beyond the critical electric field, the effect reverses and stronger momentum relaxation suppresses pair production. This demonstrates that the role of translational symmetry breaking depends sensitively on the strength of the external electric field. The analysis therefore extends the canonical framework of \cite{Sato2013} by incorporating explicit momentum-relaxing backgrounds and examining their quantitative impact across different electric field regimes.

Recent advances, such as the work \cite{Su2025}, have shown that Schwinger pair production can be significantly enhanced through modifications of the quantum phases accumulated by virtual particles, even without increasing the background electric field strength. This phase-engineering mechanism, analogous to the Aharonov--Bohm effect, demonstrates that vacuum instability may be controlled through changes in the vacuum phase structure itself. In our holographic setup, the parameter $\alpha$, which characterizes translational symmetry breaking and momentum relaxation, alters the effective holographic potential barrier and consequently changes the tunneling probability. In the subcritical regime this effect enhances tunneling by lowering the barrier, whereas in the critical and supercritical regimes it suppresses pair production. From this perspective, the $\alpha$-dependent behavior observed in the holographic model provides a qualitative analogue of phase-modified vacuum instability discussed in \cite{Su2025}, connecting translational symmetry breaking backgrounds with recent developments in quantum vacuum manipulation.

\subsection{Vacuum instability in the presence of magnetic Field}
In this subsection, we consider the total potential Eq.~\eqref{vtotfinalaby} in the presence of an external magnetic field and examine its impact on the Schwinger effect.\\
In this subsection, we distinguish between the magnetic-field components $B_{\parallel}$ and $B_{\perp}$, which are parallel and perpendicular to the external electric field, respectively. The magnetic field is normalized in units of $T_F r_0^2$, so that $B_{\parallel}/(T_F r_0^2)$ and $B_{\perp}/(T_F r_0^2)$ are dimensionless. Unless otherwise stated, the magnetic-field configurations considered in the numerical analysis have $B_{\parallel}=0$ and nonzero $B_{\perp}$. All magnetic-field comparisons are performed at fixed $\beta=E/E_c$, so the physical electric field $E$ changes when the magnetic field modifies $E_c$.

\begin{figure}[h!]
\begin{center}$
\begin{array}{cccc}
\includegraphics[width=10cm]{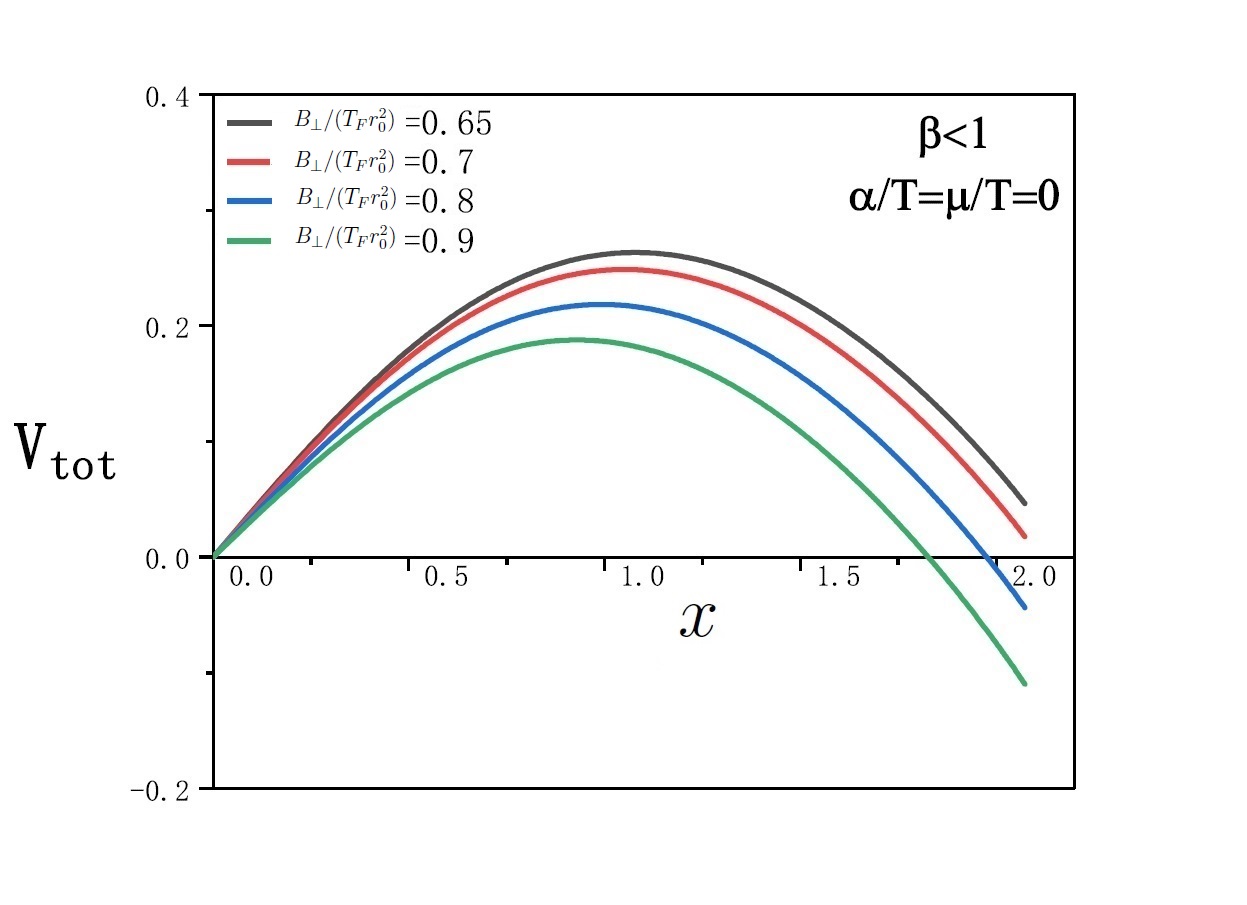}
\end{array}$
\end{center}
\caption{Total potential $V_{\mathrm{tot}}$ versus the inter-pair distance $x$ for different values of the perpendicular magnetic field $B_{\perp}$ in the subcritical regime $\beta=0.8$. The parameters are fixed to $\alpha/T=\mu/T=0$ and $B_{\parallel}=0$. The curves correspond to $B_{\perp}/(T_F r_0^2)=0.65$, $0.7$, $0.8$, and $0.9$.}
\label{VtotxB}
\end{figure}
Figure~\ref{VtotxB} shows the behavior of the total potential as the perpendicular magnetic field increases at fixed $\beta=0.8$. As $B_{\perp}$ increases, both the height and width of the potential barrier decrease, indicating a reduced tunneling barrier. Since $\beta=E/E_c$ is held fixed, the applied electric field changes simultaneously with $E_c$. Thus, the observed reduction of the potential barrier characterizes the magnetic-field dependence at fixed $\beta$, rather than at fixed physical electric field. 
This behavior indicates that the magnetic field facilitates the tunneling process by weakening the potential barrier against vacuum decay. For sufficiently large magnetic field, the system approaches the unstable regime more rapidly, implying that the magnetic field enhances pair production, also it is consistent with the reduction of the effective critical electric field in the presence of magnetic field, indicating that stronger magnetic field enhances vacuum instability and promotes the Schwinger effect \cite{zhou1912,satoyo1303,Bolo1210,Koji1403,Wen2506}.
\begin{figure}[h!]
\begin{center}$
\begin{array}{cccc}
\includegraphics[width=8cm,height=6cm,clip]{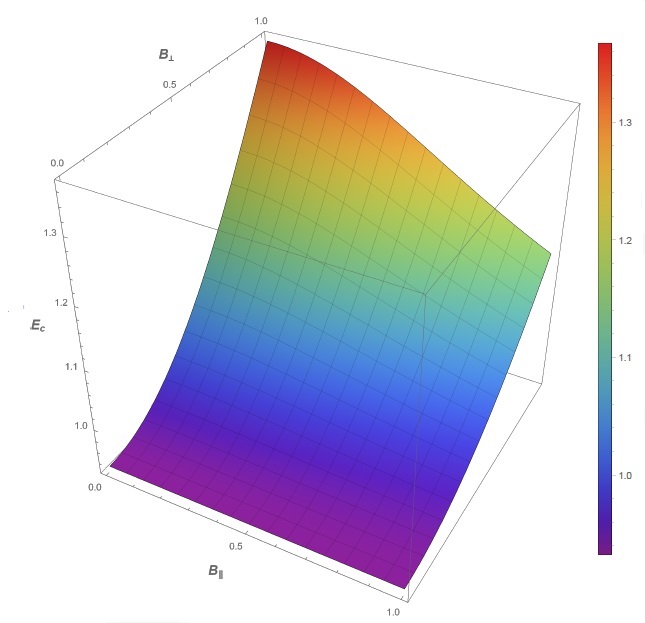}
\end{array}$
\end{center}
\caption{Critical electric field $E_c$ as a function of the perpendicular magnetic field $B_{\perp}$ and parallel magnetic field $B_{\parallel}$. The surface illustrates the combined influence of the magnetic field components on the critical electric field in the holographic background. }
\label{EcB}
\end{figure}

Figure~\ref{EcB} shows that a purely parallel magnetic field does not change the critical electric field, whereas a perpendicular magnetic field increases $E_c$. At fixed nonzero $B_{\perp}$, increasing $B_{\parallel}$ decreases $E_c$, approaching the zero-magnetic-field value as $B_{\parallel}$ becomes large. Thus, the two magnetic-field components affect the critical electric field differently, as directly follows from Eq.~\eqref{EcrB}.

Therefore, the two components of the magnetic field affect the critical behavior differently, and their competition determines the effective threshold for vacuum instability in the holographic system. This analysis also clarifies how the external magnetic field enters the Schwinger mechanism without modifying the background geometry itself.

\subsubsection{Subcritical Regime $\beta<1 $}
\begin{figure}[h!]
\begin{minipage}[c]{1\textwidth}
\tiny{(a)}\includegraphics[width=8cm,height=5cm,clip]{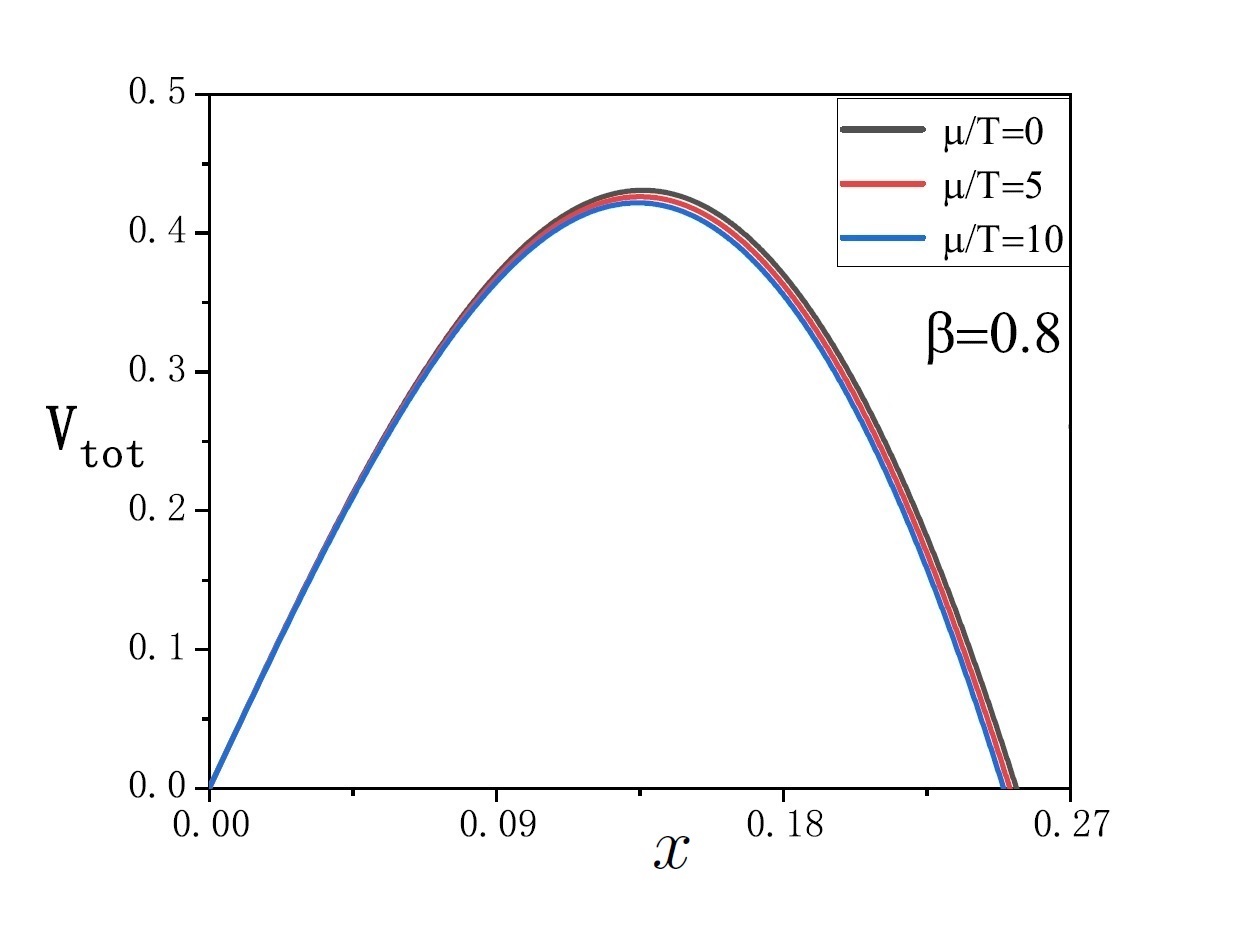}
\hspace{0.2cm}
\tiny{(b)}\includegraphics[width=8cm,height=5cm,clip]{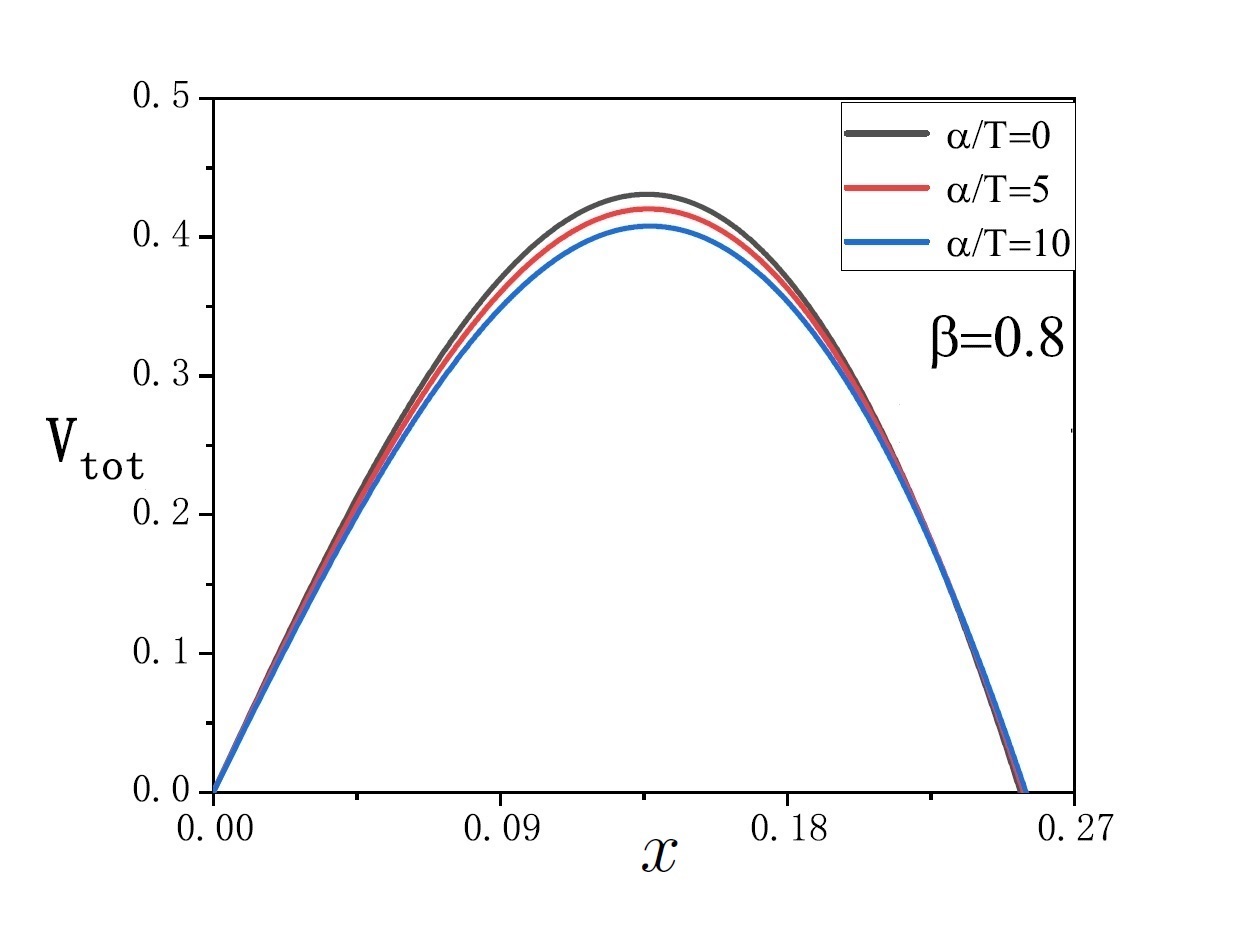}
\end{minipage}
\caption{Total potential $V_{\mathrm{tot}}$ versus the inter-pair distance $x$ in the presence of magnetic field ($B_{\perp}/(T_F r_0^2)=0.74$, $B_{\parallel}=0$) for the subcritical regime $\beta=0.8$. The parameters are fixed as $r_0=5$ and $b=0.4$ ($r_h=2$). In panel (a), $\alpha/T=0$ and $\mu/T$ varies, whereas in panel (b), $\mu/T=0$ and $\alpha/T$ varies.}
\label{7}
\end{figure}
Figure~\ref{7} illustrates the total potential in the presence of magnetic field for the subcritical regime $(\beta<1)$. In panel (a), increasing $\mu/T$ lowers the potential barrier and slightly decreases the turning point of the potential. Therefore, finite chemical potential continues to facilitate the tunneling process in the presence of magnetic field.
Similarly, panel (b) shows that increasing $\alpha/T$ also suppresses the potential barrier. Hence, the qualitative effect of momentum relaxation in the subcritical phase remains unchanged after switching on the magnetic field.

Comparing Figs.~\ref{tunneling} and ~\ref{7}, the inclusion of magnetic field changes the magnitude of the potential while preserving the qualitative behavior induced by both $\mu/T$ and $\alpha/T$ in the subcritical regime.
\subsubsection{Critical Regime $\beta=1$}
Figure.~\ref{8} presents the total potential at the critical electric field in the presence of magnetic field. In panel (a), increasing $\mu/T$ shifts the potential downward, indicating that finite chemical potential continues to support the onset of Schwinger pair production near the critical point.
In contrast, panel (b) shows that increasing $\alpha/T$ shifts the potential upward, as highlighted in the zoomed region. Therefore, momentum relaxation suppresses Schwinger pair production in the critical regime even in the presence of magnetic field.

The comparison between panels (a) and (b) demonstrates that the opposite roles played by chemical potential and momentum relaxation persist after including the magnetic field. This also confirms that the qualitative change in the role of momentum relaxation near the critical electric field is robust against magnetic effects.
\begin{figure}[h!]
\begin{minipage}[c]{1\textwidth}
\tiny{(a)}\includegraphics[width=8cm,height=5cm,clip]{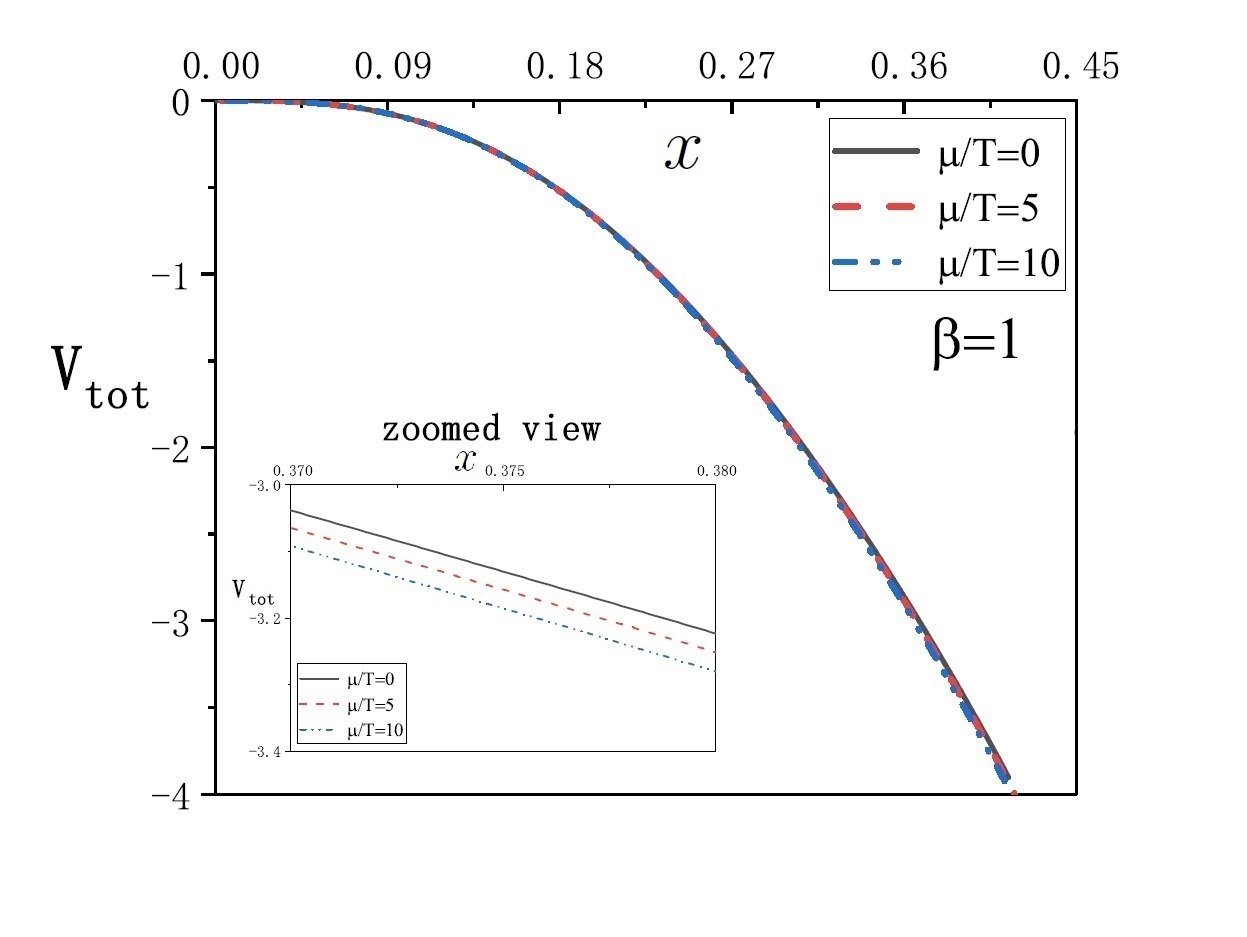}
\hspace{0.2cm}
\tiny{(b)}\includegraphics[width=8cm,height=5cm,clip]{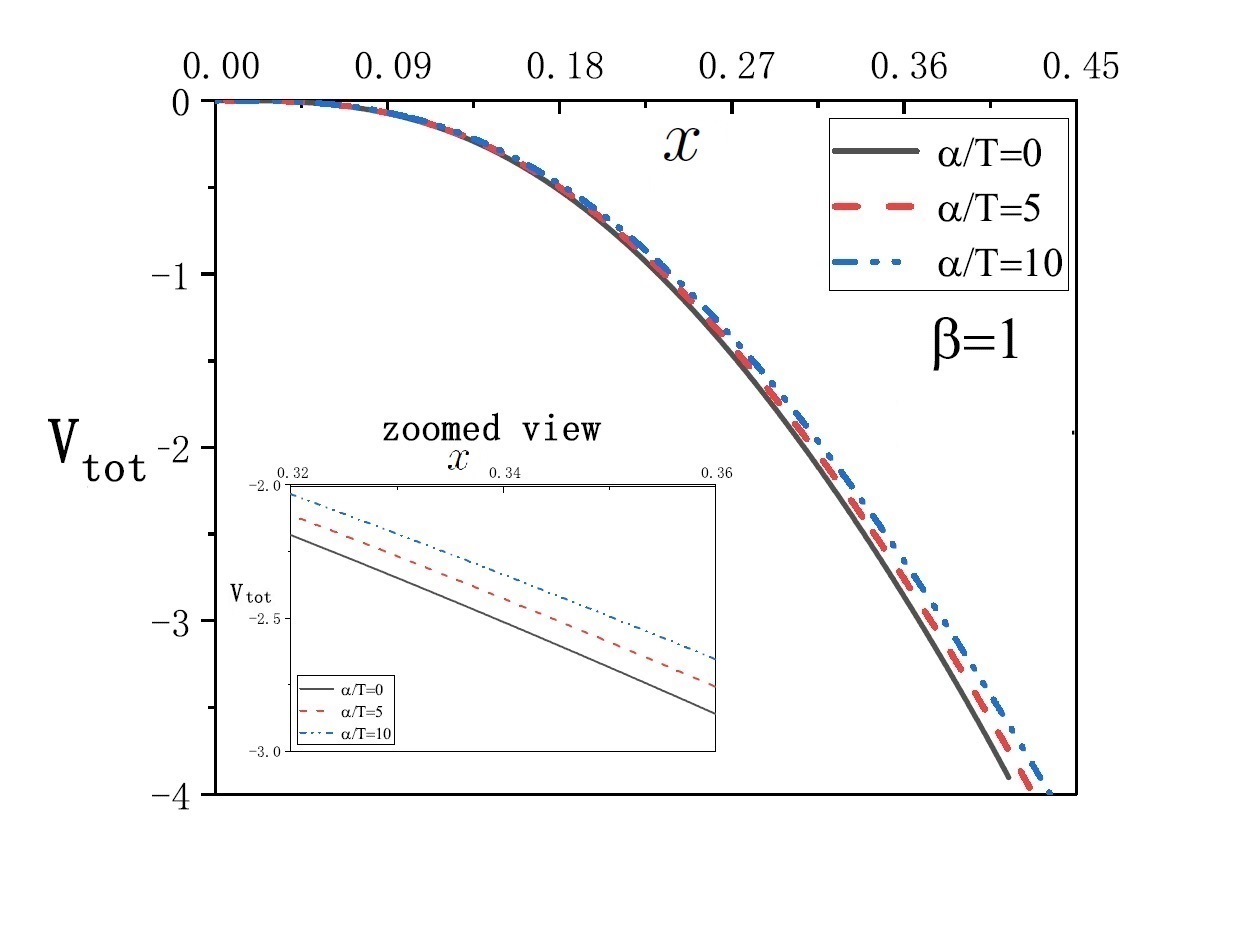}
\end{minipage}
\caption{Total potential $V_{\mathrm{tot}}$ versus the inter-pair distance $x$ in the presence of magnetic field ($B_{\perp}/(T_F r_0^2)=0.74$, $B_{\parallel}=0$)  for the critical regime $\beta=1$. The parameters are chosen as $r_0=5$ and $b=0.4$ ($r_h=2$). In panel (a), $\alpha/T=0$ and $\mu/T$ varies, while in panel (b), $\mu/T=0$ and $\alpha/T$ varies.}
\label{8}
\end{figure}
\subsubsection{Supercritical  Regime $\beta>1 $}

Figure~\ref{9} shows the behavior of the total potential in the supercritical regime in the presence of magnetic field. In panel (a), increasing $\mu/T$ shifts the potential toward lower values and strengthens the instability associated with Schwinger pair production.

Conversely, panel (b) indicates that increasing $\alpha/T$ shifts the potential upward, leading to a comparatively weaker instability. This behavior becomes more apparent in the zoomed region.

Comparing the critical and supercritical regimes, the qualitative influence of $\mu/T$ and $\alpha/T$ remains unchanged after including the magnetic field: finite chemical potential favors pair production, whereas momentum relaxation suppresses it. This again demonstrates that momentum relaxation changes its qualitative role once the system approaches and crosses the critical electric field.\\

\begin{figure}[h!]
\begin{minipage}[c]{1\textwidth}
\tiny{(a)}\includegraphics[width=8cm,height=5cm,clip]{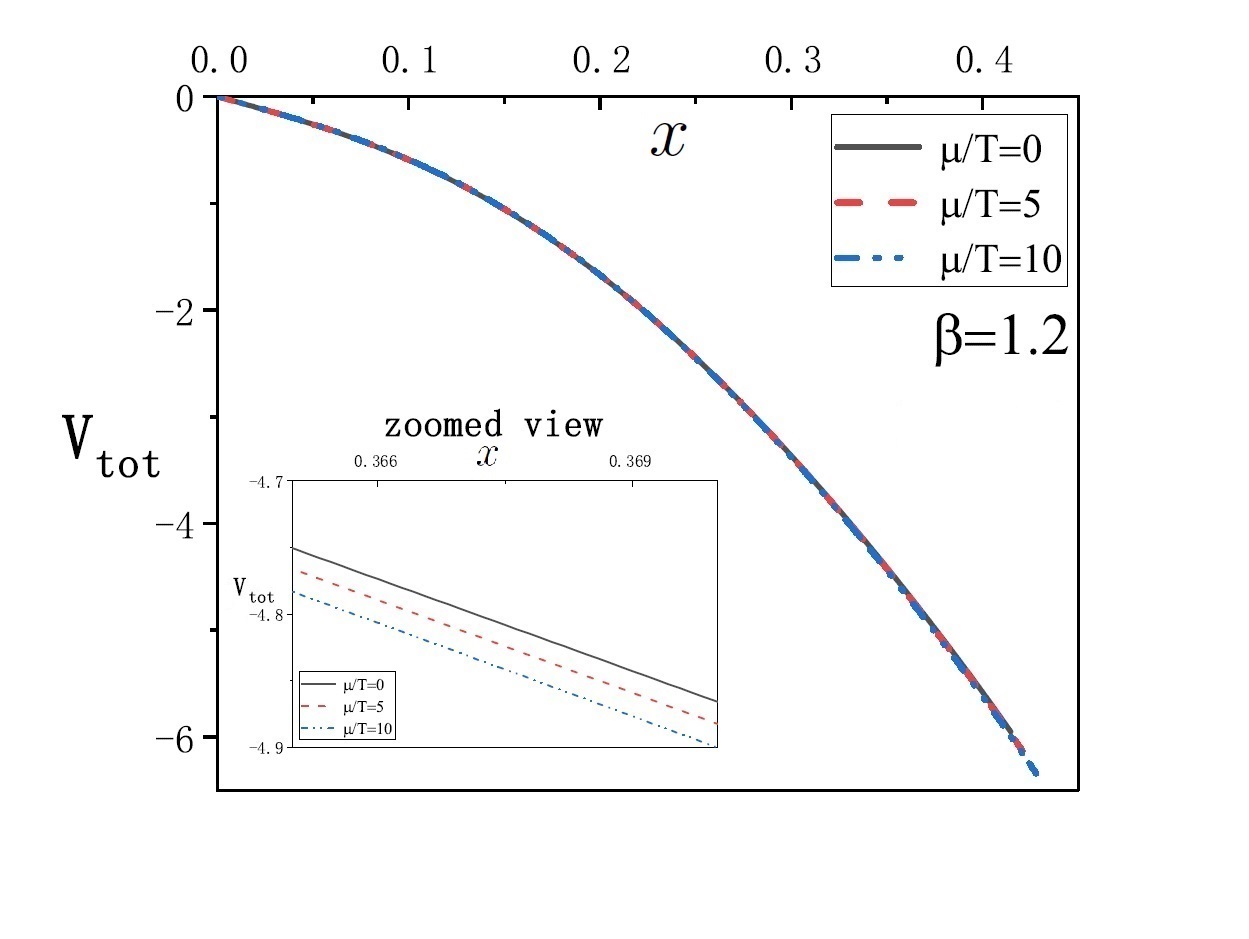}
\hspace{0.2cm}
\tiny{(b)}\includegraphics[width=8cm,height=5cm,clip]{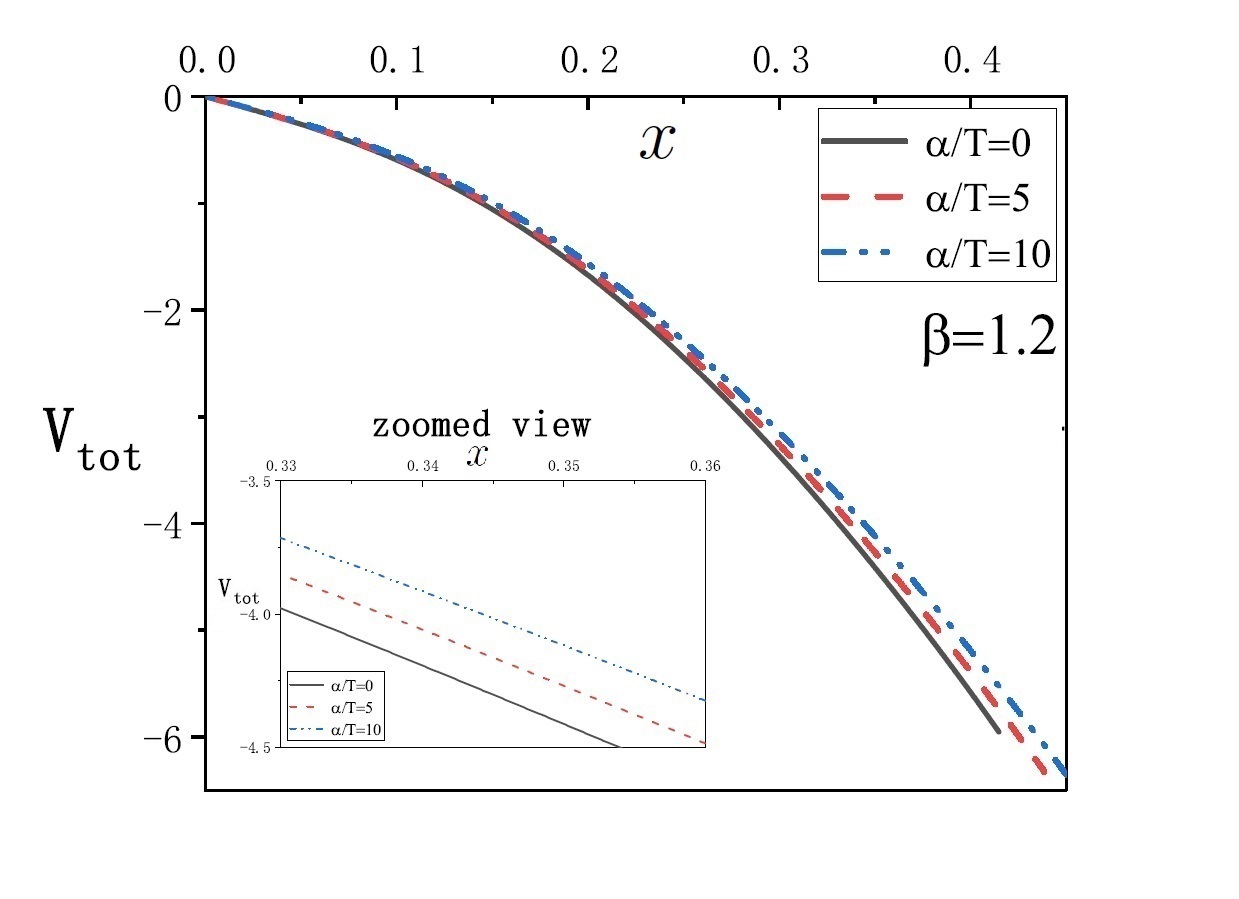}
\end{minipage}
\caption{Total potential $V_{\mathrm{tot}}$ versus the inter-pair distance $x$ in the presence of magnetic field ($B_{\perp}/(T_F r_0^2)=0.74$, $B_{\parallel}=0$) for the supercritical regime $\beta=1.2$. The parameters are fixed as $r_0=5$ and $b=0.4$ ($r_h=2$). In panel (a), $\alpha/T=0$ and $\mu/T$ varies, whereas in panel (b), $\mu/T=0$ and $\alpha/T$ varies.}
\label{9}
\end{figure}
\section{Qualitative proxy for the Schwinger pair production rate in the TSB background}\label{se:Gamma}

The Schwinger pair production rate can be estimated semiclassically from the Euclidean worldsheet action associated with the classical string configuration. Using the Nambu--Goto action introduced in Eq.~\eqref{Lgeneral} together with the classical string profile obtained from Eq.~\eqref{detadx}, the corresponding on-shell Euclidean action can be written as,
\begin{equation}\label{Scl}
S_{\mathrm{cl}}=2\,\mathcal{T}\,T_F\int_{r_c}^{r_0}dr\sqrt{\frac{f(r)\,r^2}{f(r)\,r^2-f(r_c)\,r_c^2}},
\end{equation}
where $\mathcal{T}$ denotes the Euclidean time interval.

The relation $\Gamma\sim e^{-S_{\mathrm{cl}}}$ in Eq.~\eqref{rate} provides the standard semiclassical connection between the pair production rate and the classical Euclidean action.
\begin{equation}\label{rate}
\Gamma\sim e^{-S_{\mathrm{cl}}}.
\end{equation}

 In the present work, however, we do not perform an independent instanton calculation of the production rate. Instead, we use the total potential to construct a qualitative proxy for the behavior of the rate, defined by 
\begin{equation}\label{effectiveRate}
\Gamma_{\mathrm{eff}}
\propto \exp\left(-V_{\mathrm{tot}}\right).
\end{equation}
This quantity is introduced solely as a qualitative proxy derived from the total potential. It is therefore intended to characterize the relative suppression or enhancement associated with changes in the potential barrier, rather than to represent an independently calculated pair production rate.

The external magnetic field considered here lives on the probe D3-brane worldvolume and does not modify the background geometry. Its effect enters through the DBI action and changes the effective critical electric field appearing in the total potential. Consequently, its influence on the qualitative proxy is mediated through the corresponding modification of the potential barrier.

\subsection{Qualitative proxy for the Schwinger pair production rate in the absence of magnetic field}
Since the Schwinger production process becomes significantly enhanced near and above the critical electric field, we focus our analysis on the critical and supercritical regimes.
\subsubsection{Critical Regime $\beta=1$}
\begin{figure}[h!]
\begin{minipage}[c]{1\textwidth}
\tiny{(a)}\includegraphics[width=8cm,height=5cm,clip]{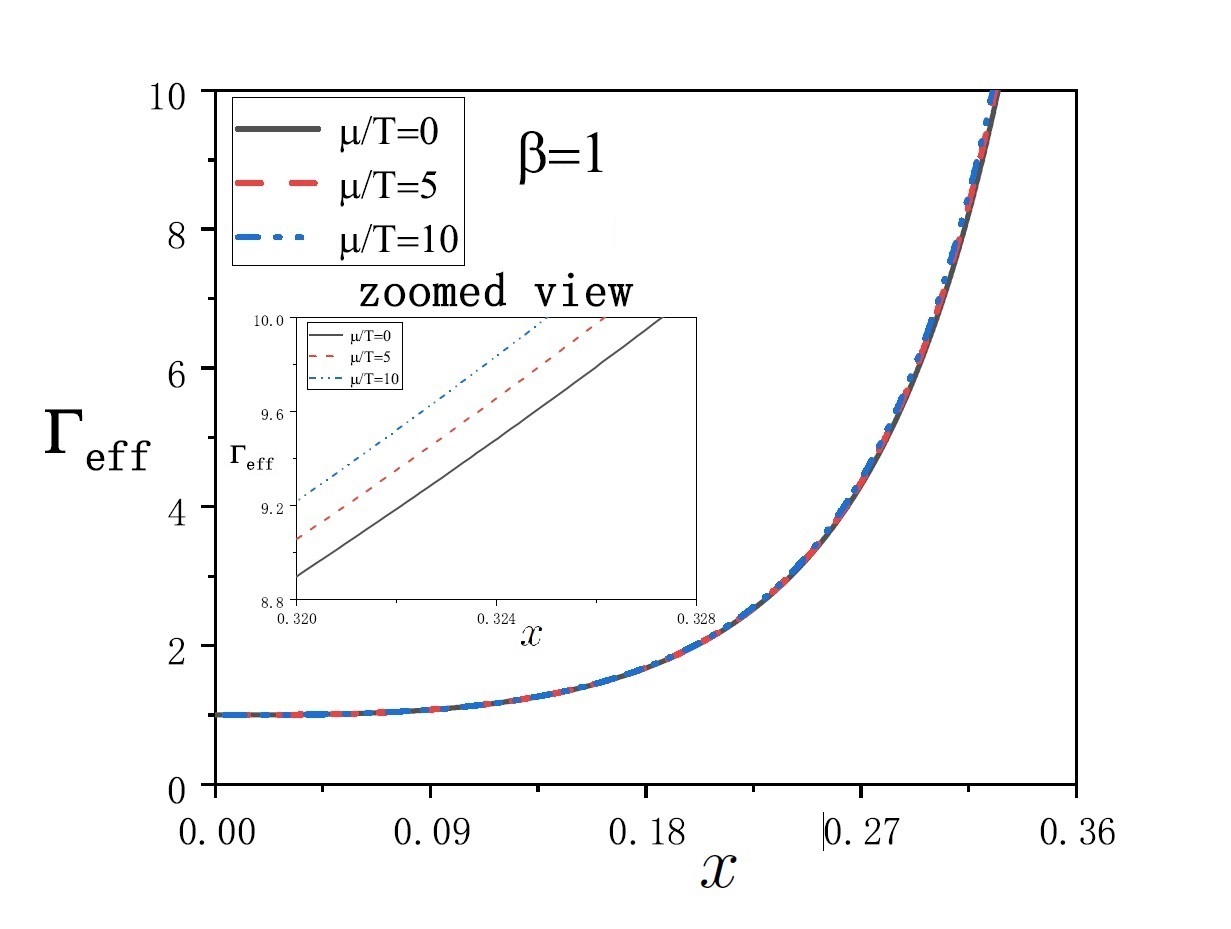}
\hspace{0.2cm}
\tiny{(b)}\includegraphics[width=8cm,height=5cm,clip]{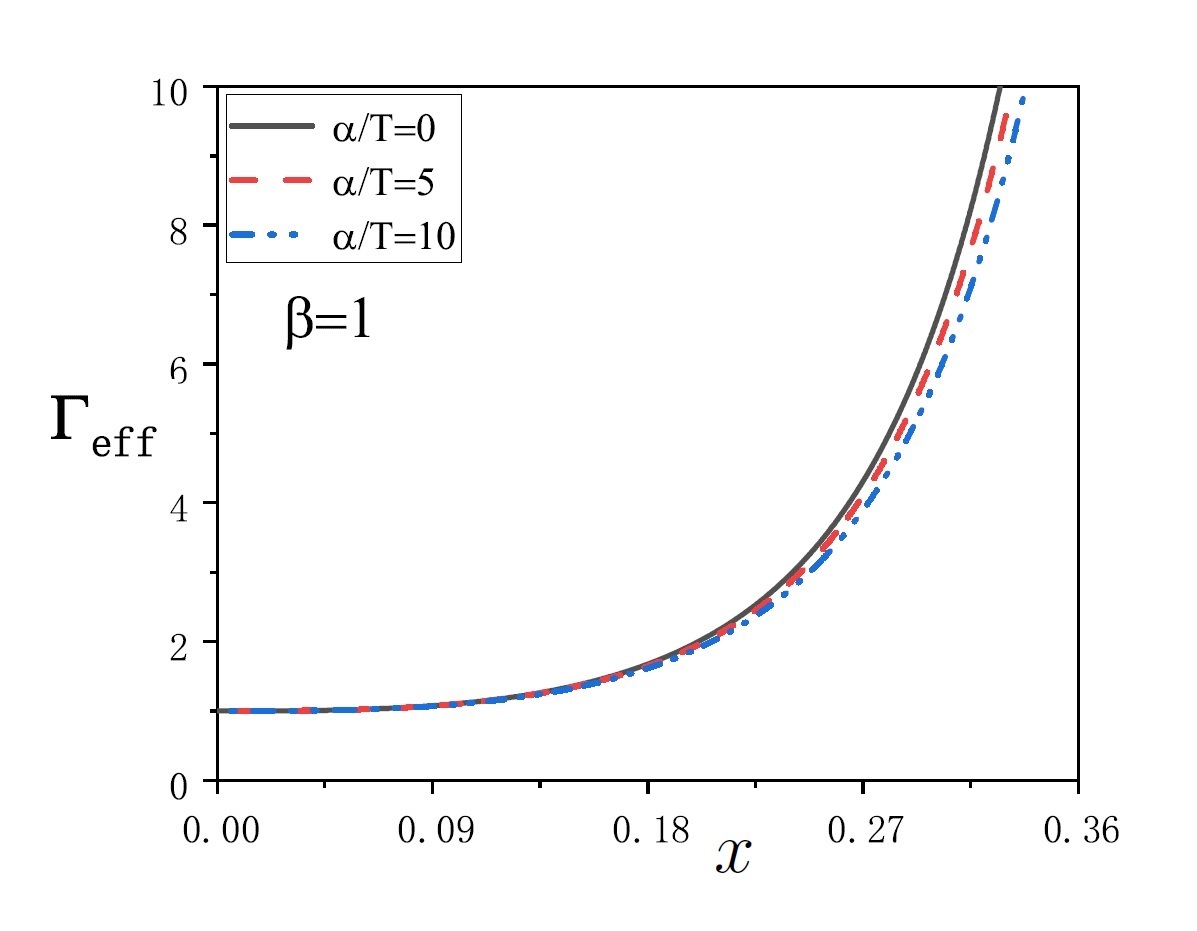}
\end{minipage}
\caption{Qualitative proxy for the Schwinger pair production rate $\Gamma_{\mathrm{eff}}$ as a function of the inter-pair separation $x$ in the absence of magnetic field ($B=0$) for the critical regime $\beta=1$. Panel (a) corresponds to $\alpha=0$ and varying $\mu/T$, while panel (b) corresponds to $\mu=0$ and varying $\alpha/T$. The inset plots provide a magnified view of the small-$x$ region. The parameters are chosen as $r_0=5$, $b=0.4$, and $r_h=2$.}
\label{rate_crit_noB}
\end{figure}

Figure~\ref{rate_crit_noB} illustrates the qualitative proxy for the Schwinger pair production rate at the critical electric field, where the potential barrier disappears. Compared with the subcritical regime, the  qualitative proxy grows much more rapidly with the inter-pair separation, indicating enhanced vacuum instability near the critical electric field.

In panel (a), increasing $\mu/T$ enhances the qualitative proxy over the entire range of $x$. This behavior follows from the corresponding total potential profiles, where larger $\mu/T$ shifts the potential downward and facilitates pair production.

In contrast, panel (b) demonstrates that increasing $\alpha/T$ suppresses the  qualitative proxy for pair production. This behavior is consistent with the corresponding critical potential profiles, where larger values of $\alpha/T$ shift the total potential upward and therefore suppress the production process.

Therefore, in the critical regime the effects of chemical potential and translational symmetry breaking become qualitatively different: finite chemical potential enhances the Schwinger production process, whereas stronger translational symmetry breaking suppresses it.
\subsubsection{Supercritical Regime $\beta>1$}
\begin{figure}[h!]
\begin{minipage}[c]{1\textwidth}
\tiny{(a)}\includegraphics[width=8cm,height=5cm,clip]{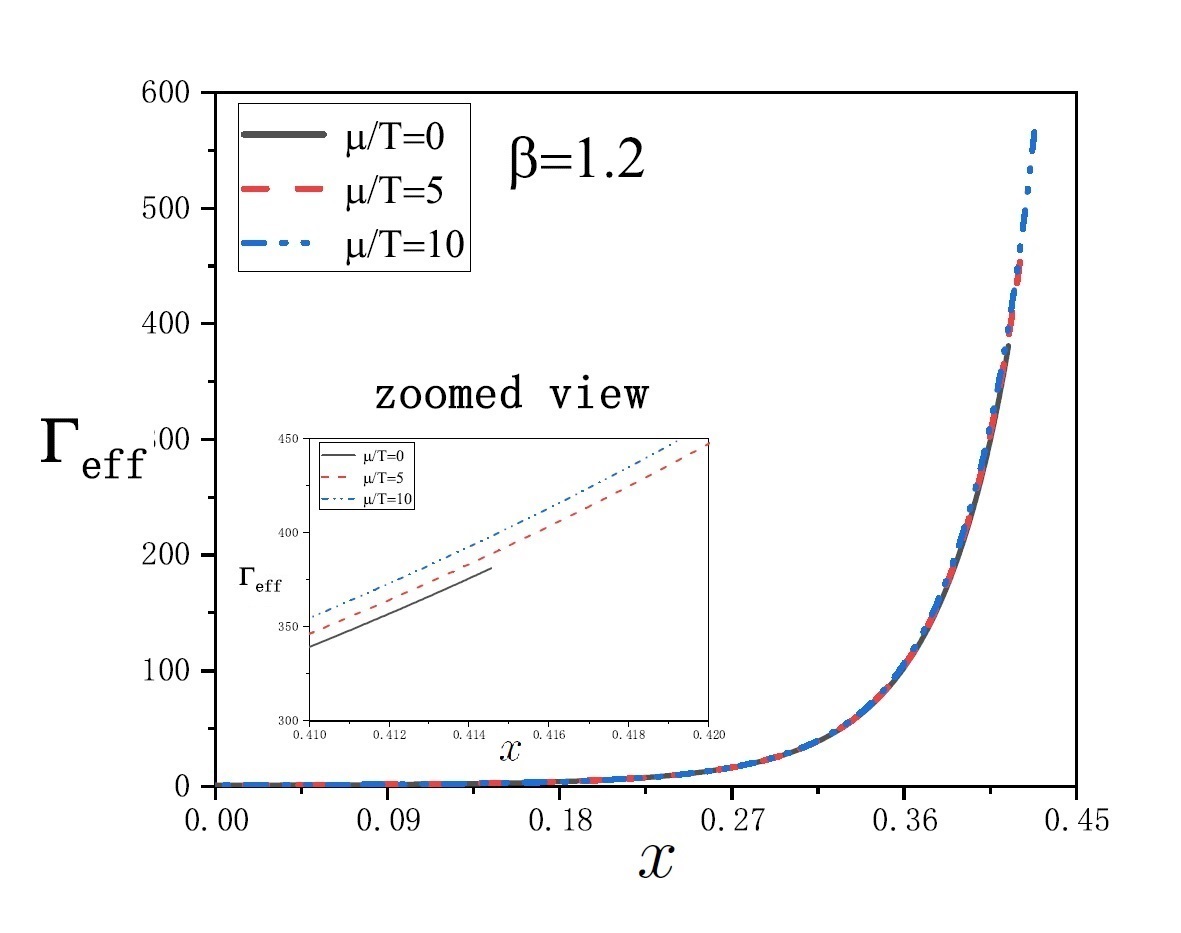}
\hspace{0.2cm}
\tiny{(b)}\includegraphics[width=8cm,height=5cm,clip]{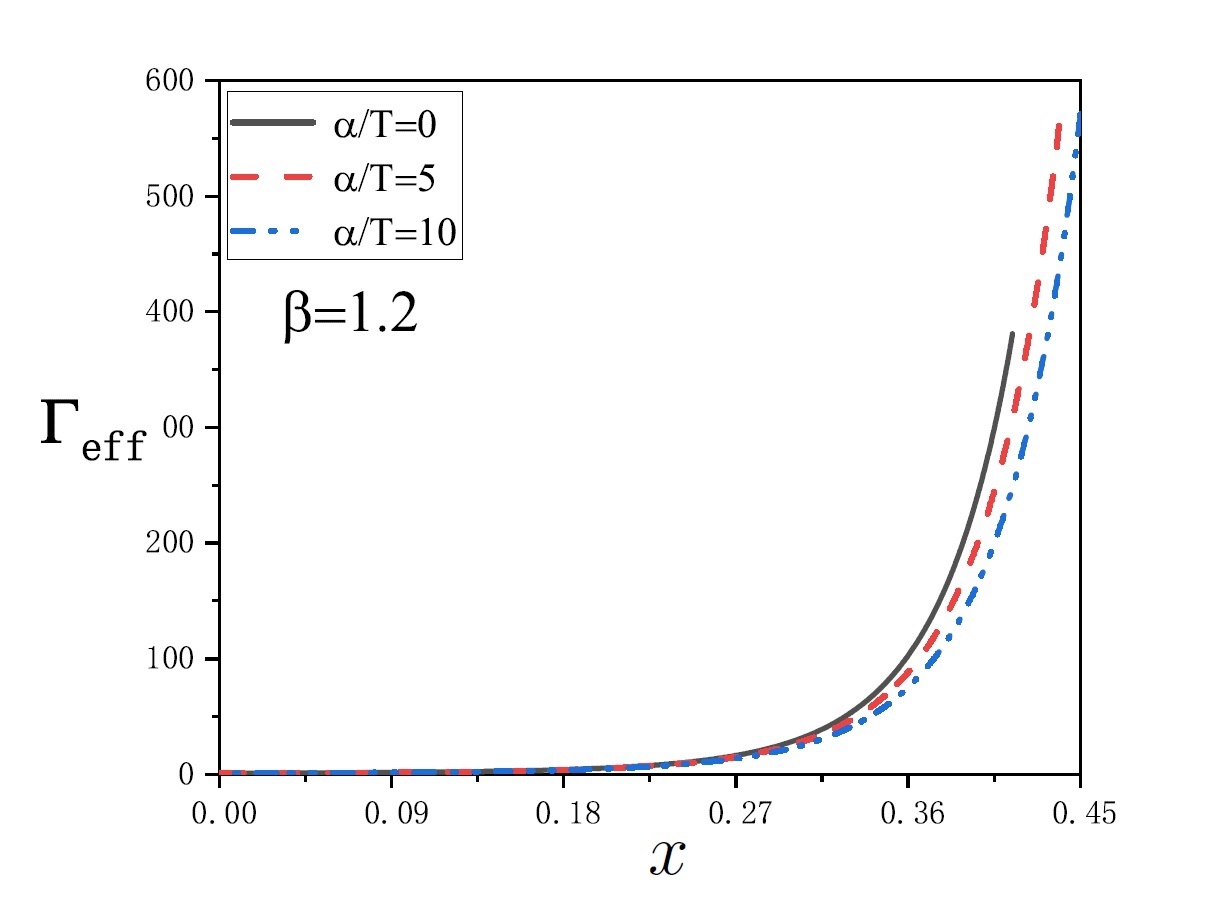}
\end{minipage}
\caption{Qualitative proxy for the Schwinger pair production rate $\Gamma_{\mathrm{eff}}$ as a function of the inter-pair separation $x$ in the absence of magnetic field ($B=0$) for the supercritical regime $\beta>1$ with $\beta=1.2$. Panel (a) corresponds to $\alpha=0$ and varying $\mu/T$, while panel (b) corresponds to $\mu=0$ and varying $\alpha/T$. The inset plots show the small-$x$ behavior. The parameters are fixed to $r_0=5$, $b=0.4$, and $r_h=2$.}
\label{rate_super_noB}
\end{figure}

Figure~\ref{rate_super_noB} presents the qualitative proxy for the Schwinger pair production rate in the supercritical regime, where the electric field exceeds the critical value and the vacuum becomes highly unstable. In this regime, the  qualitative proxy increases extremely rapidly with the inter-pair separation, indicating enhanced pair production.

Panel (a) shows that increasing $\mu/T$ further enhances the  qualitative proxy. This result is consistent with the corresponding total potential behavior, where larger chemical potential lowers the total potential profile and facilitates the production process.

Panel (b) demonstrates that increasing $\alpha/T$ suppresses the qualitative proxy in the supercritical regime. The upward shift of the total potential induced by larger $\alpha/T$ suppresses the production process compared with the lower-$\alpha/T$ configurations.

Comparing the critical and supercritical regimes, one observes that the qualitative roles of $\mu/T$ and $\alpha/T$ remain unchanged, although the overall enhancement of the Schwinger production process becomes significantly stronger for $\beta>1$.

\subsection{Qualitative proxy for the Schwinger pair production rate in the presence of magnetic field}
\subsubsection{Critical Regime $\beta=1$}
\begin{figure}[h!]
\begin{minipage}[c]{1\textwidth}
\tiny{(a)}\includegraphics[width=8cm,height=5cm,clip]{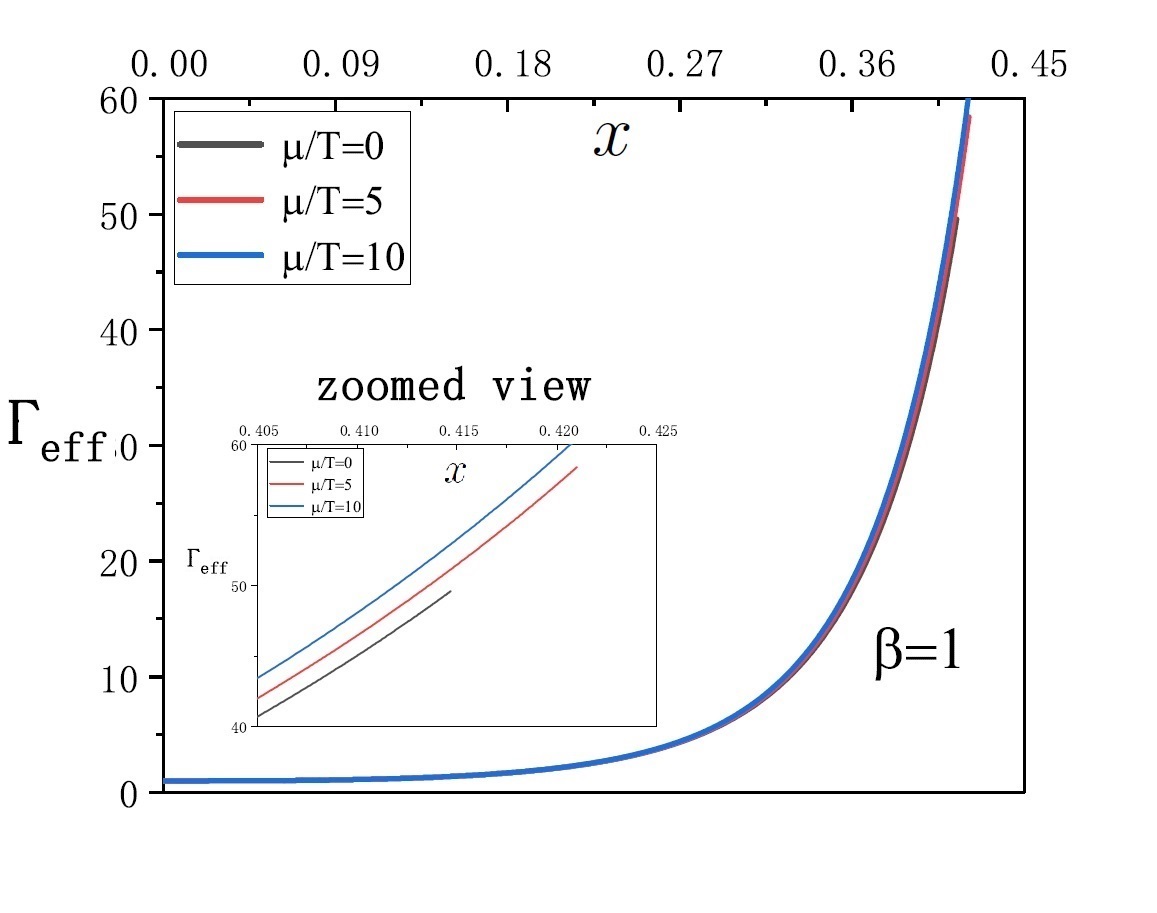}
\hspace{0.2cm}
\tiny{(b)}\includegraphics[width=8cm,height=5cm,clip]{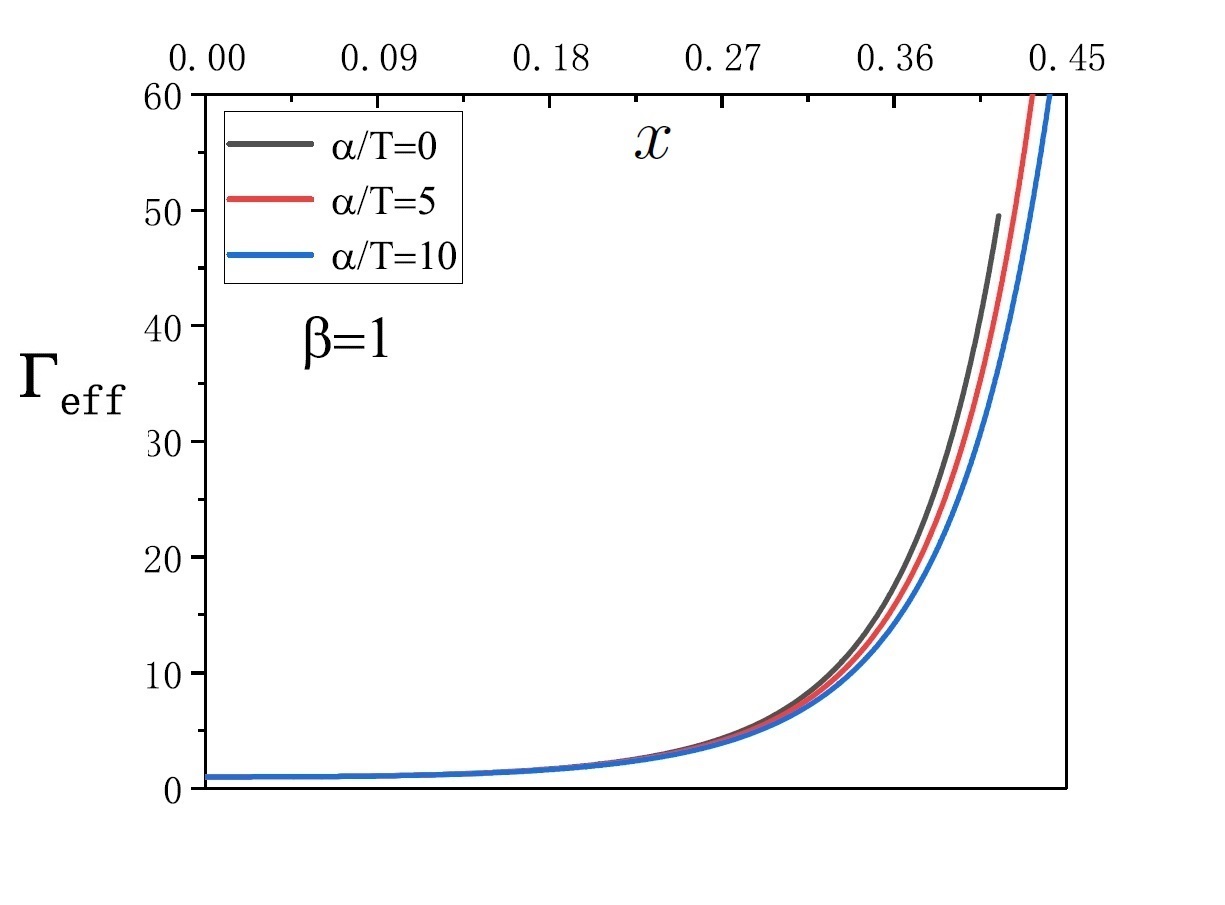}
\end{minipage}
\caption{Qualitative proxy for the Schwinger pair production rate $\Gamma_{\mathrm{eff}}$ as a function of the inter-pair separation $x$ in the presence of magnetic field ($B_{\perp}/(T_F r_0^2)=0.74$, $B_{\parallel}=0$) for the critical regime $\beta=1$. Panel (a) corresponds to $\alpha=0$ and varying $\mu/T$, while panel (b) corresponds to $\mu=0$ and varying $\alpha/T$. The inset plots provide a magnified view of the small-$x$ region. The parameters are fixed to $r_0=5$, $b=0.4$, and $r_h=2$.}
\label{rate_crit_B}
\end{figure}
All magnetic-field comparisons in this subsection are performed at fixed $\beta=E/E_c$. Therefore, the physical electric field $E$ changes when the magnetic field modifies $E_c$.

Figure~\ref{rate_crit_B} illustrates the qualitative proxy for Schwinger pair production rate at the critical electric field in the presence of magnetic field. The  qualitative proxy increases rapidly with the inter-pair separation, indicating enhanced vacuum instability near the critical electric field.

In panel (a), increasing $\mu/T$ enhances the qualitative proxy, in agreement with the corresponding critical potential behavior. The downward shift of the total potential facilitates the production process.

On the other hand, panel (b) shows that increasing $\alpha/T$ suppresses the qualitative proxy. This behavior is consistent with the corresponding total potential analysis, where larger values of $\alpha/T$ shift the potential upward in the critical regime.

Therefore, even in the presence of magnetic field, the chemical potential and translational symmetry breaking parameter affect the Schwinger production process differently in the critical regime: finite density enhances the production process, whereas stronger translational symmetry breaking suppresses it.
\subsubsection{Supercritical Regime $\beta>1$}
\begin{figure}[h!]
\begin{minipage}[c]{1\textwidth}
\tiny{(a)}\includegraphics[width=8cm,height=5cm,clip]{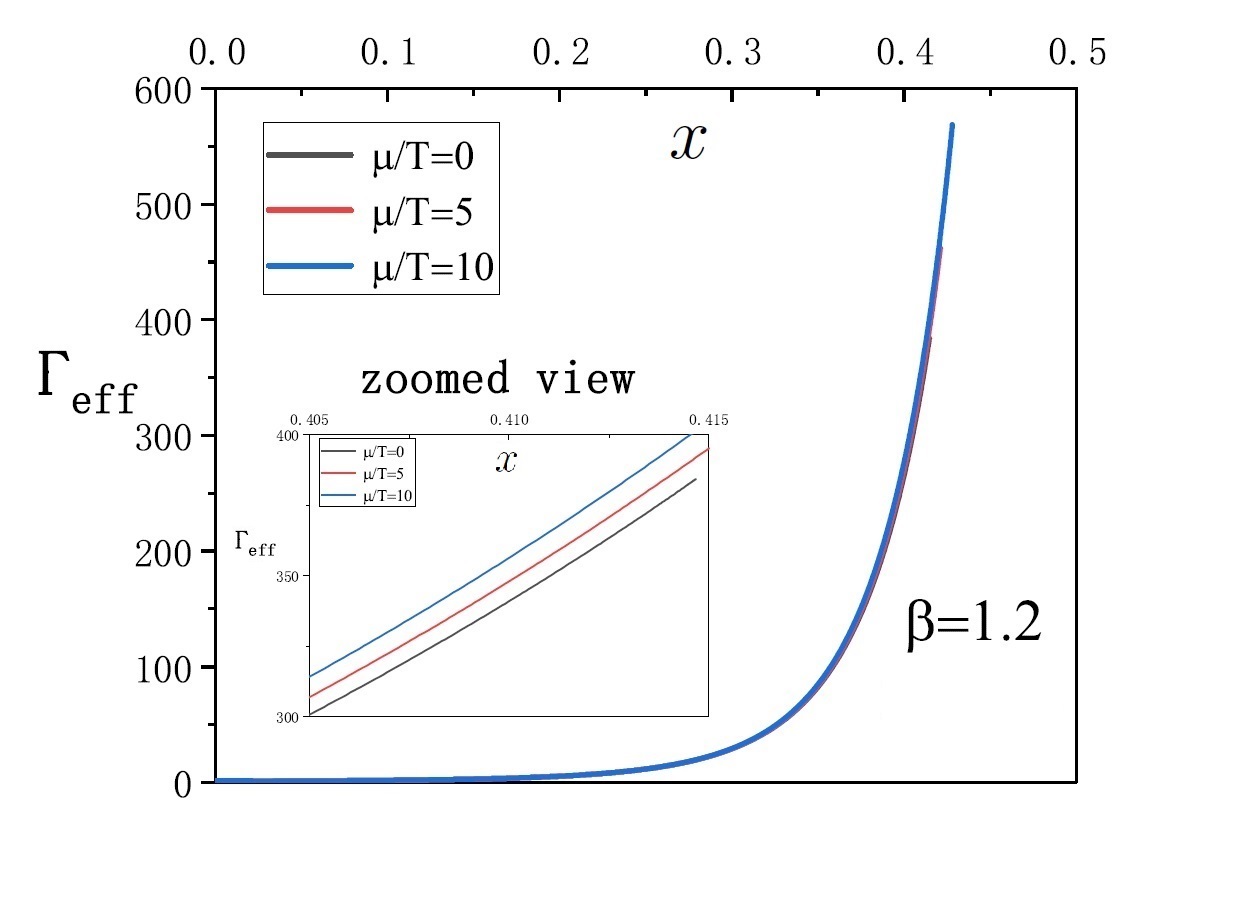}
\hspace{0.2cm}
\tiny{(b)}\includegraphics[width=8cm,height=5cm,clip]{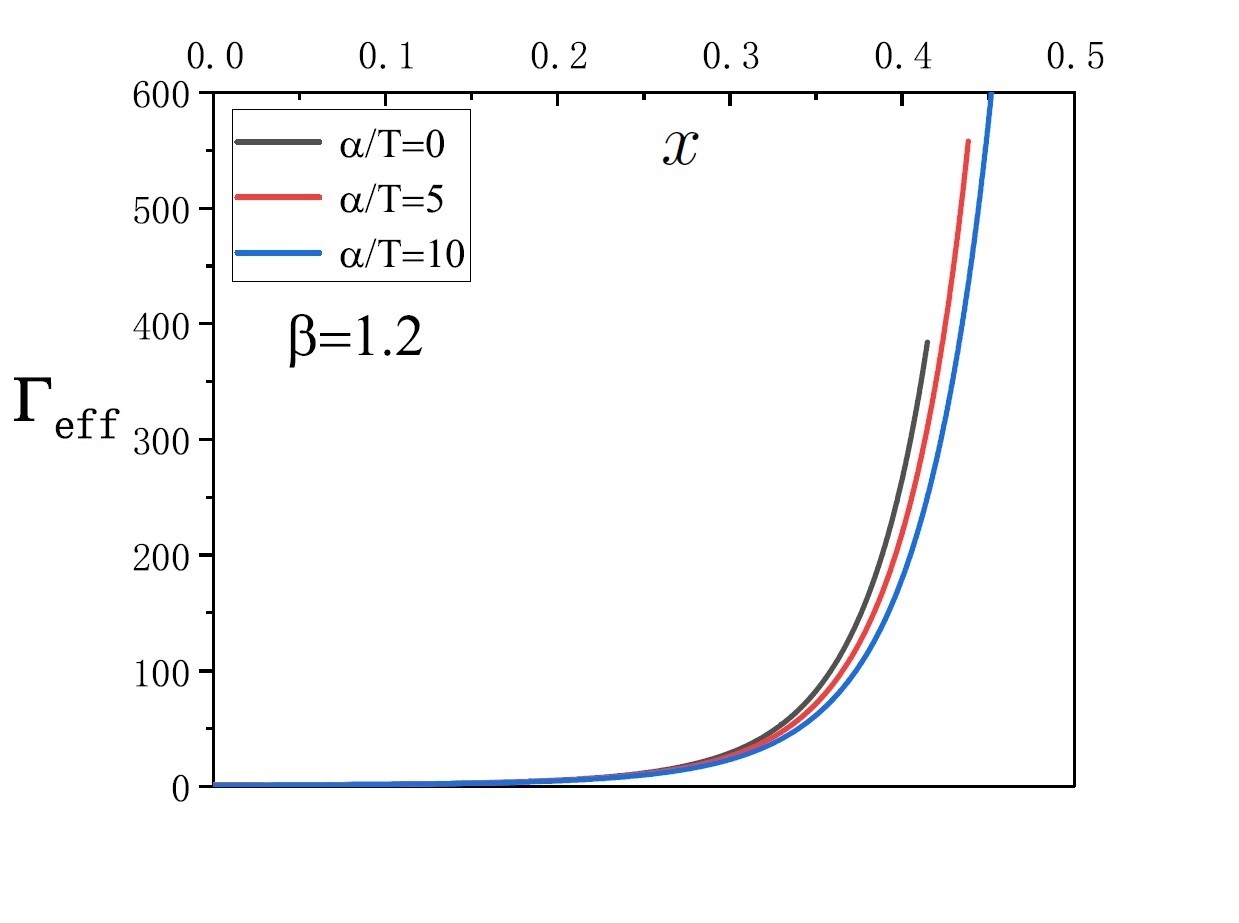}
\end{minipage}
\caption{Qualitative proxy for the Schwinger pair production rate $\Gamma_{\mathrm{eff}}$ as a function of the inter-pair separation $x$ in the presence of magnetic field ($B_{\perp}/(T_F r_0^2)=0.74$, $B_{\parallel}=0$) for the supercritical regime $\beta>1$ with $\beta=1.2$. Panel (a) corresponds to $\alpha=0$ and varying $\mu/T$, while panel (b) corresponds to $\mu=0$ and varying $\alpha/T$. The inset plots show the small-$x$ behavior. The parameters are chosen as $r_0=5$, $b=0.4$, and $r_h=2$.}
\label{rate_super_B}
\end{figure}

Figure~\ref{rate_super_B} presents the qualitative proxy for  Schwinger pair production rate in the supercritical regime in the presence of magnetic field. In this regime,  the qualitative proxy grows extremely rapidly, signaling rapid vacuum decay for electric fields larger than the critical value.

Panel (a) shows that increasing $\mu/T$ enhances the qualitative proxy, consistent with the corresponding total potential behavior. The chemical potential lowers the effective potential profile and consequently facilitates the Schwinger production process.

In panel (b), increasing $\alpha/T$ suppresses the qualitative proxy. This behavior again agrees with the total potential analysis, where larger values of $\alpha/T$ shift the potential upward in the supercritical regime and therefore reduce the growth of the  qualitative proxy for Schwinger pair production rate.

Comparing the magnetic and non-magnetic cases, the qualitative roles of $\mu/T$ and $\alpha/T$ remain unchanged, while the magnetic field quantitatively modifies the production behavior through its contribution to the effective critical electric field.

\begin{figure}[h!]
\begin{center}$
\begin{array}{cccc}
\includegraphics[width=10cm]{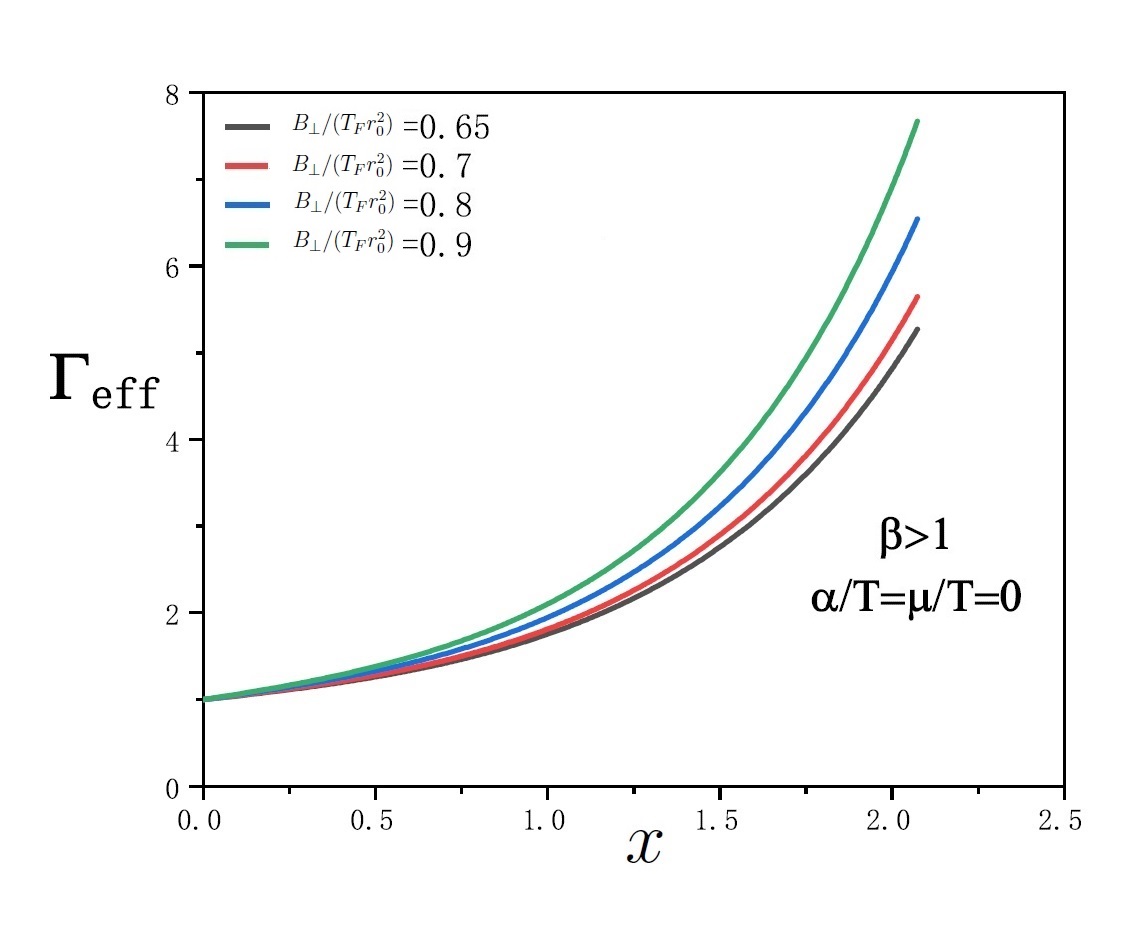}
\end{array}$
\end{center}
\caption{Qualitative proxy for the Schwinger pair production rate $\Gamma_{\mathrm{eff}}$ as a function of the inter-pair separation $x$ for different values of the magnetic field parameters $B_{\perp}/(T_F r_0^2)=0.65, 0.7, 0.8,0.9$ and $B_{\parallel}=0$ in the supercritical regime $\beta>1$. The parameters are fixed to $\alpha/T=\mu/T=0$. The numerical calculations are performed with $r_0=5$, $b=0.4$, and $r_h=2$. }
\label{rate_B_super}
\end{figure}

Figure~\ref{rate_B_super} illustrates the dependence of the qualitative proxy on the perpendicular magnetic field in the supercritical regime at fixed $\beta=1.2$. Increasing $B_{\perp}$ lowers the total potential barrier and consequently increases the qualitative proxy defined in Eq. (41). This behavior reflects the corresponding change in the total potential at fixed $\beta$, rather than an independently calculated change in the physical pair production rate.
\begin{figure}[h!]
\begin{center}$
\begin{array}{cccc}
\includegraphics[width=10cm]{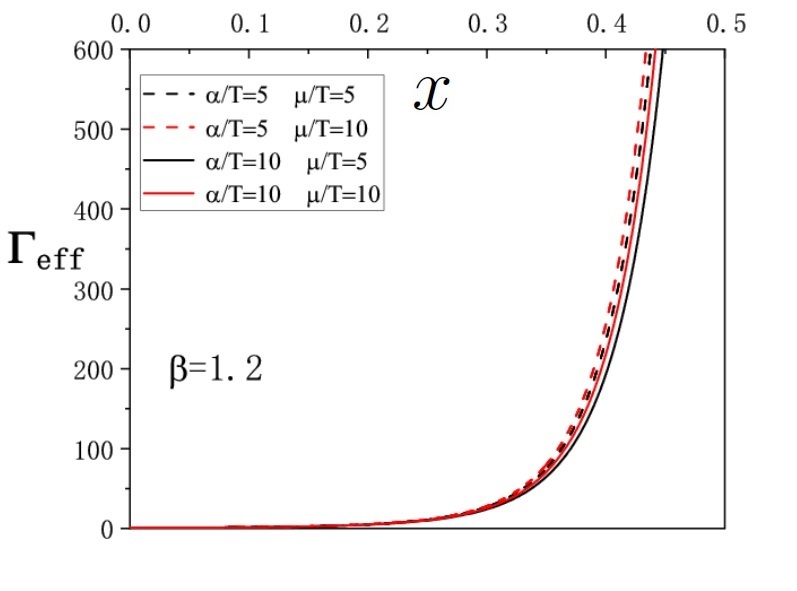}
\end{array}$
\end{center}
\caption{Qualitative proxy for the Schwinger  pair production rate $\Gamma_{\mathrm{eff}}$ as a function of the inter-particle separation $x$ in the supercritical regime $\beta=1.2$ in the presence of magnetic field ($B_{\perp}/(T_F r_0^2)=0.74$, $B_{\parallel}=0$). The dashed curves correspond to $\alpha/T=5$, while the solid curves correspond to $\alpha/T=10$. 
The black curves represent $\mu/T=5$, whereas the red curves represent $\mu/T=10$. 
The parameters are fixed as $r_0=5$, $b=0.4$, and $r_h=2$.}
\label{gammamualpha}
\end{figure}

Figure~\ref{gammamualpha} illustrates the combined influence of the chemical potential and translational symmetry breaking on the qualitative proxy for Schwinger pair production rate in the supercritical regime. 
For fixed $\alpha/T$, increasing $\mu/T$ enhances the qualitative proxy, as indicated by the upward shift from the black curves to the red curves. 
In contrast, for fixed $\mu/T$, increasing $\alpha/T$ suppresses the qualitative proxy, as seen from the downward shift of the solid curves relative to the dashed ones.

The figure therefore demonstrates the competition between finite density and translational symmetry breaking in the holographic Schwinger effect. 
While the chemical potential facilitates pair production by reducing the effective suppression, stronger momentum relaxation acts oppositely and suppresses the production process. 
This confirms that the effects of $\mu/T$ and $\alpha/T$ remain qualitatively distinct even when both parameters are simultaneously nonzero.

\section{Summary and Outlook}\label{Summ}

In this work, we investigated the holographic Schwinger effect in a background with translational symmetry breaking (TSB) and finite chemical potential. The geometry is characterized by two independent physical parameters: the TSB parameter \(\alpha\), which controls momentum relaxation in the dual field theory, and the chemical potential \(\mu\), which determines the finite density of the system. Using the potential analysis method based on the evaluation of Wilson loops on the probe brane, we derived the total potential for a virtual quark--antiquark pair and analyzed the onset of vacuum instability and pair production. The critical electric field \(E_{c}\) emerges as the threshold at which the potential barrier disappears.

Our analysis demonstrates that the effects of \(\alpha\) and \(\mu\) on the Schwinger process are qualitatively different and strongly depend on the dynamical regime. In the subcritical regime, increasing either \(\alpha\) or \(\mu\) reduces the potential barrier and facilitates the pair production process. However, near and above the critical electric field, the roles of these two parameters become distinct. While increasing the chemical potential continuously lowers the total potential and enhances the Schwinger pair production process, increasing the translational symmetry breaking parameter shifts the potential upward and suppresses the production process. These results clearly show that translational symmetry breaking and finite density influence the non-perturbative vacuum instability in fundamentally different ways.

We further extended the analysis by introducing an external magnetic field. The external magnetic field enters through the DBI action and modifies the effective critical electric field without altering the background geometry. For the configurations considered here, with $B_{\parallel}=0$, increasing the perpendicular magnetic field lowers the potential barrier at fixed $\beta=E/E_c$, while the physical electric field changes through the corresponding variation of $E_c$.

We also introduced a qualitative proxy for the Schwinger pair production rate through its exponential dependence on the total potential. This proxy is directly derived from the potential analysis and is used only to characterize the relative suppression or enhancement associated with changes in the potential barrier; it should not be interpreted as an independently calculated production rate.

Overall, our study reveals the rich interplay among translational symmetry breaking, finite density, and external magnetic field in the holographic Schwinger effect. The interplay among translational symmetry breaking, finite density, and the magnetic-field dependence of the critical electric field provides a unified picture of non-perturbative pair production in strongly coupled systems.

As possible future directions, it would be interesting to investigate the holographic Schwinger effect in the presence of time-dependent electric fields, anisotropic backgrounds, or higher-derivative corrections. Extending the analysis to more general holographic setups may provide further insights into vacuum instability and non-perturbative phenomena in strongly correlated quantum systems.\\
\\
\textbf{\Large{Acknowledgement}}\\

The work is supported by the National Natural Science Foundation of China (Grant Nos. 1267050014 and 12275067), Science and Technology Innovation Leading Talent Support Program of Henan Province  (Grant No. 254200510039) , Science and Technology R$\&$D Program Joint Fund Project of Henan Province  (Grant No.225200810030), Henan Province Foreign Scientists Studio (Grant No. GZS2020032), and National Key R$\&$D Program of China (Grant No. 2023YFA1606000).

\end{document}